\documentclass{article}
\usepackage[hypertexnames=false]{hyperref}       
\usepackage{arxiv}

\usepackage[utf8]{inputenc} 
\usepackage[T1]{fontenc}    
\usepackage{url}            
\usepackage{booktabs}       
\usepackage{amsfonts}       
\usepackage{nicefrac}       
\usepackage{microtype}      
\usepackage{lipsum}
\usepackage{amsmath}
\usepackage{graphicx}
\usepackage[superscript]{cite}
\usepackage{authblk}
\usepackage[pagewise]{lineno}

\graphicspath{{Images/}}

\title{\textbf{Neuromorphic metamaterials for mechanosensing and perceptual associative learning}}

\author[1,+]{Katherine S. Riley}
\author[2,+]{Subhadeep Koner}
\author[1,+]{Juan C. Osorio}
\author[2]{Yongchao Yu}
\author[1]{Harith Morgan}
\author[1]{Janav P. Udani}
\author[2,*]{Stephen A. Sarles}
\author[1,*]{Andres F. Arrieta}
\affil[1]{School of Mechanical Engineering, Purdue University, West Lafayette, IN 47907 USA}
\affil[2]{Department of Mechanical, Aerospace and Biomedical Engineering, University of Tennessee, Knoxville, TN 37996 USA}
\affil[+]{authors contributed equally}
\affil[*]{corresponding authors: aarrieta@purdue.edu; ssarles@utk.edu}

\begin{document}
\maketitle
\begin{abstract} \normalsize

Physical systems exhibiting neuromechanical functions promise to enable structures with directly encoded autonomy and intelligence. We report on a class of neuromorphic metamaterials embodying bioinspired mechanosensing, memory, and learning functionalities obtained by leveraging mechanical instabilities and flexible memristive materials. Our prototype system comprises a multistable metamaterial whose bistable units filter, amplify, and transduce external mechanical inputs over large areas into simple electrical signals using piezoresistivity. We record these mechanically transduced signals using non-volatile flexible memristors that remember sequences of mechanical inputs, providing a means to store spatially distributed mechanical signals in measurable material states. The accumulated memristance changes resulting from the sequential mechanical inputs allow us to physically encode a Hopfield network into our neuromorphic metamaterials. This physical network learns a series of external spatially distributed input patterns. Crucially, the learned patterns input into our neuromorphic metamaterials can be retrieved from the final accumulated state of our memristors. Therefore, our system exhibits the ability to learn without supervised training and retain spatially distributed inputs with minimal external overhead. Our system's embodied mechanosensing, memory, and learning capabilities establish an avenue for synthetic neuromorphic metamaterials enabling the learning of touch-like sensations covering large areas for robotics, autonomous systems, wearables, and morphing structures.

\end{abstract}


\vfill

The nervous systems of animals comprise networks of distributed sensory, memory, and control elements that enable perception, reaction and adaptation in response to varied external stimuli. The coevolution of the nervous system and body morphology is thought to reduce the complexity of sensed signals due to morphological computing and short neuronal connections. This results in a form of neuromechanical control that requires few inputs to fulfil complex functions.~\cite{Laschi2016b,Kim2013a} Some organisms, such as comb jellies (phylum Ctenophora~\cite{Moroz2015}), perform complex tasks even without a central nervous system. This capability stems from coevolved neuromechanical systems that exploit morphological and sensory couplings to create autonomic responses,~\cite{Kim2013} which constitute one of the simplest forms of learned behavior. Typical autonomic behaviors include reflexes,~\cite{Proctor2010} preflexes, and central pattern generators,~\cite{Futakata2011} all of which leverage fast, decentralized sense-compute-actuate control loops.~\cite{Laschi2016a} At the most basic level, this is achieved through the synergy of morphology, sensing, computation, and actuation systems that perceive stimuli,~\cite{Fratzl2009} filter noise,~\cite{Mcconney2007} and store valuable information in the physical state of the systems. Achieving similar capabilities in synthetic systems is important for realizing the next generation of autonomic, multifunctional, power-efficient materials, devices, and systems. Single artificial neuromechanical functions encoded into material systems have been demonstrated, including sensing~\cite{Jung2018,Wang2021}, filtering and thresholding~\cite{Park2015,LeFerrand2018}, and different forms of mechanically-encoded memory~\cite{Keim2019,Chen2021}.

Learned autonomic biological responses are thought to be created when groups of neurons reinforce synaptic connections, resulting in simultaneous firing.~\cite{Jain1996,Fields2020} Inspired by synapses and neurons, neuromorphic computing has emerged as a field leveraging history-dependent responses and electrical tunability of materials and devices to approximate brain-inspired computation.~\cite{Xia2019a} Many neuromorphic systems exploit memory resistive, or memristive, materials to achieve the characteristics of synaptic plasticity, memory storage, and even threshold-gated neural spiking. Thus, memristors offer a low-power, analog route for highly parallelized computation.~\cite{Tang2019}. 
Recently, neuromorphic materials have been integrated with sensory systems to encode and process external inputs that enable the formation of memory and learning\cite{Park2020}.
In a seminal work, afferent nervous functions have been mimicked via neuromorphic circuits to convert information from mechanical inputs into motor functions, thereby completing a synthetic reflex arc.~\cite{Kim2018}. Artificial neuromechanical perception has also been demonstrated in flexible substrates embedding synaptic transistors \cite{Wan2018} and Nafion-based memristors \cite{Zhang219} capable of durably encoding spatiotemporal tactile excitations into signals amenable for \textit{ex-situ} classification using supervised learning. Spatiotemporal, long-term memory formation and retention has also been shown using a 3x3 array of CNT/PDMS piezoresistive sensors and Pt/HfO$_2$/TiN memristors.~\cite{Xia2021}

\begin{figure}[!t]
  \centering
  \includegraphics[width=\textwidth]{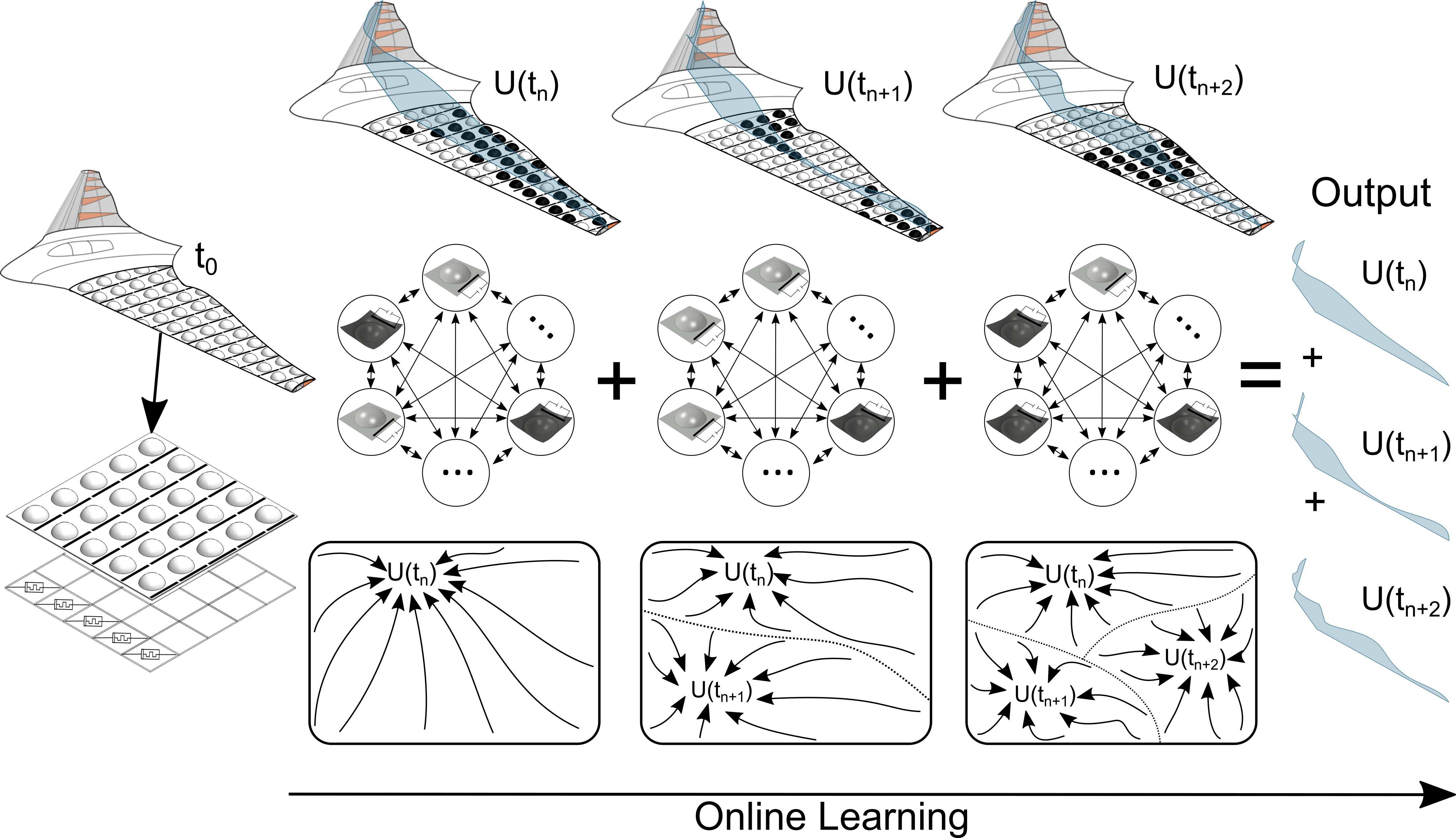}
  \caption{Neuromorphic metamaterial composed of mechanosensing unit cells with inherent filtering, transduction, mechanoelectric signal amplification, and memory capabilities. The units' transduction drives non-volatile memristors that display cumulative memristance variation when excited above a voltage threshold mediated by the dome state-dependent piezoresistive sensor. These characteristics allow for realizing physical Hopfield networks capable of encoding several spatiotemporal patterns as attractors of the network's interaction matrix, the synaptic weights of which are stored as the final (current) states of the memristors. The physical Hopfield networks enabled by our neuromorphic metamaterials are trained online by continuously-changing distributed mechanical inputs (such as pressures over wings). This allows for realizing substrates exhibiting event camera-like tactile perception~\cite{Taunyazov2020} with online learning and reduced external memory as several patterns can be retrieved from the memristors' final states.}
  \label{fi:conceptual}
\end{figure}

Online learning (implying adaptation as data is sensed) is a crucial function of neuromechanical systems that has so far been much less explored in mechanically reconfigurable, neuromorphic material systems. Thus, synthetic material systems enabling integrated mechanosensing, memory, and autonomic learning capabilities remain rare. Realizing synthetic systems that embody natural neuromechanical systems' functions promises to enable structures with advanced functionalities such as autonomy and intelligence. We report on a class of neuromorphic metamaterials embodying bioinspired mechanosensing, memory, and learning functionalities achieved by leveraging local mechanical instabilities and soft memristive materials. Our prototype consists of a multistable metamaterial whose mechanosensing units filter, amplify, and transduce into simple electrical signals external mechanical inputs applied over large areas (c.a. 10$^2$ cm$^2$) using threshold-dependent structural bistability and embedded piezoresistive sensors (Figure \ref{fi:conceptual}). We encode the transduced signals using non-volatile flexible memristors, thus providing a means to store spatially distributed mechanical signals in analog material states. Notably, the additive nature of the accumulated memristance changes allows us to physically encode into our neuromorphic metamaterials a Hopfield network~\cite{Hopfield1982,Hopfield1985} that learns a series of applied spatially distributed mechanical inputs. This is achieved with memristor arrays connected to capture pair-wise interactions between different mechanosensing units yielding the Hopfield network's connectivity matrix. Crucially, the learned, mechanically applied patterns can be retrieved from the final electrical resistance states of the memristors. Therefore, our system displays the ability to learn online, retaining spatially distributed inputs exceeding a critical threshold with minimal external overhead. 

The combined mechanosensing and thresholding behavior of our neuromorphic metamaterial is reminiscent of event cameras~\cite{Lichtsteiner2006,Gallego2022} in which pixels asynchronously record events when light intensity exceeds a specified threshold, offering power-efficient and low-latency visual perception. In such systems, pixel triggering events are timestamped and recorded in silicon-implemented memory components. Our metamaterial provides a tactile analog to an event camera where recording occurs when the mechanical input exceeds a threshold value; however, it is augmented with the capacity to sequentially encode these input events as memories in the form of accumulated final memristance states. In the sense that the Hopfield network is continuously updated as tactile events exceed a threshold, our neuromorphic metamaterial offers a new route for realizing engineering systems with online learning through tactile perception for applications in robotics, autonomous systems, wearables, and morphing structures. 

\begin{figure}[!t]
  \centering
  \includegraphics[width=\textwidth]{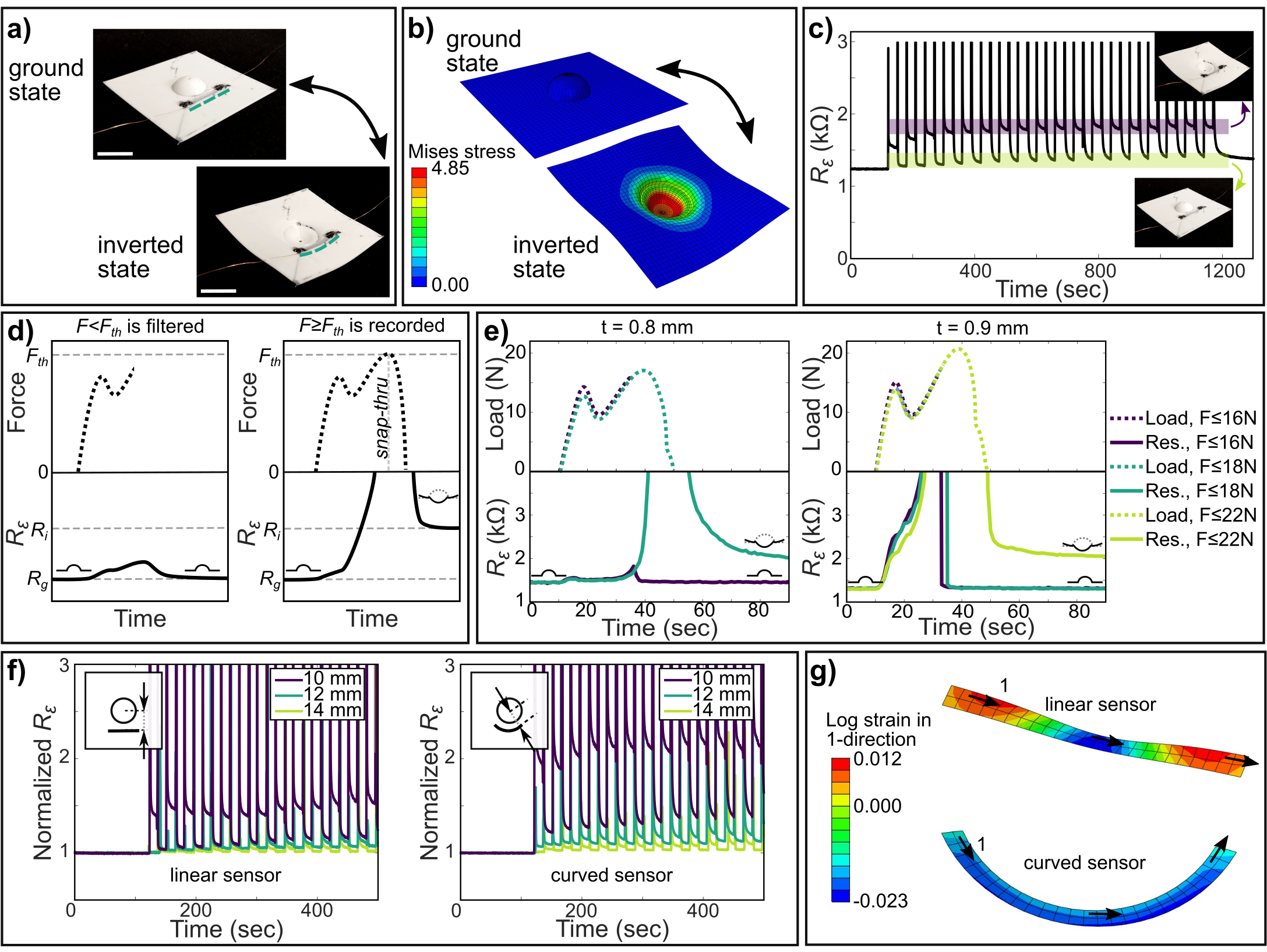}
  \caption{a) The bistable unit cell in its ground and inverted states. The sensor is flat in the ground state and curves out-of-plane in the inverted state. Scale bars are 16 mm, the diameter of the dome. b) FEA stress plots of the unit cell dome without an embedded sensor. c) The resistance of the sensor, $R_{\epsilon}$, over 18 inversion cycles with 30 second intervals between snap-through events. After the initial increase in resistance, the unit cell displays two distinct, consistent levels of resistance corresponding to the stable states. The dome has dimensions $h$ = 7 mm, $t$ = 0.8 mm, and $r_d$ = 8 mm. d) The bistability thresholding functionality is schematically illustrated. Forces below the threshold, $F<F_{th}$, do not induce snap-through and are filtered out: the sensor's initial and final resistances correspond to the ground state, $R_g$. Forces at or above the threshold, $F\geq F_{th}$ cause the dome to undergo snap-through to the inverted state, and the event is recorded via the sensor's final resistance corresponding to the inverted state, $R_i$. e) The filtering tunability is experimentally demonstrated by the changing snap-through load for unit cells of two different thicknesses, $t$ = 0.8 mm and $t$ = 0.9 mm ($h$ = 7 mm, $r_d$ = 8 mm).  f) Linear and curved sensors at distances of 10, 12, and 14 mm from the center of the dome ($h$ = 7 mm, $t$ = 0.8 mm, $r_d$ = 8 mm) are compared to demonstrate changing nonlinear signal amplification. g) Isolated FEA strain plots for the linear and curved sensors at 10 mm from the center of the dome in the inverted state.}
  \label{fi:mechanosensing}
\end{figure}

\textit{Mechanosensing--} The mechanosensing unit cell is comprised of a bistable, compliant dome with a piezoresistive sensor embedded in the flat membrane surrounding the dome. This membrane may connect multiple dome unit cells to form a metasheet. The geometry of the unit cell is defined by the dome height, $h$; thickness, $t$; and radius, $r_d$; as well as the base width, $w$, and piezoresistive sensor dimensions (Figure \ref{fi:domeschematic}) \cite{Faber2020}. Each dome-shaped unit is geometrically bistable with two stable states: the stress-free, as-printed \textit{ground} state and the \textit{inverted} state (Figure \ref{fi:mechanosensing}a). The state of the dome determines the deflection and strain in the surrounding base membrane (Figure \ref{fi:mechanosensing}b). The bistability of the dome consequently means the sensor at the base also exhibits two stable configurations with corresponding resistance levels. Thus, the electrical resistance of the piezoresistive sensor, $R_{\epsilon}$, and the structural behavior of the dome are intrinsically linked. 

We demonstrate this mechanoelectrical coupling over multiple inversion cycles using 3D printed unit cells composed of TPU with conductive PLA sensors (Figure \ref{fi:mechanosensing}c). The initial cycles differ from the steady-state largely due to the strain softening behavior of the TPU substrate, which is most significant in the first inversion cycles (Figure \ref{fi:domecycle1to10}) \cite{Christ2017}. This effect is compounded by permanent damage induced by strain in the stiffer and more brittle conductive PLA sensor.~\cite{Sau2000}. Thus, repeatedly strained sensors show higher ground state resistance than the as-printed resistance \cite{Christ2017,Liu2016a,Sau2000,Schouten2020}. This baseline drift can be reduced by inverting the domes several times before use. After this initial phase, the unit cell has two distinct, repeatable resistance levels (Figure \ref{fi:mechanosensing}c), with the lower resistance level, $R_{\epsilon}=R_g$, indicating the ground state and the higher resistance level, $R_{\epsilon}=R_i$, indicating the inverted state. The difference in resistance between the two states is on the order of several hundred Ohms. The short transient spikes in between steady-state values are due to the high strains experienced during dome snap-through between states, which cause the sensors to briefly lose conductivity \cite{Christ2017,LeFerrand2019a}.

\textit{Filtering and Thresholding--} Filtering is a critical function of neuromechanical systems that allows them to consolidate external stimuli to those which which meet a particular threshold. In our neuromorphic metasheet, switching between the stable states of the dome is a threshold-dependent process: a minimum force, $F_{th}$, is required to trigger snap-through of the dome from the ground state to the inverted state. The force and energy required for snap-through are proportional to the material properties and geometry as follows, where $E$ is the elastic modulus and $\nu$ is the Poisson’s ratio \cite{Faber2020,Seffen2016}:

\begin{equation}
    F_{th} \sim \textit{Inversion energy} \sim \frac{Et^3}{1-\nu^2}  \left( \frac{4h^2}{r_d^2} \right)
    \label{eq:invenergy}
\end{equation}

Forces below the inversion threshold, $F<F_{th}$, do not cause the dome to snap-through to the inverted state. Therefore, when a sub-threshold force is removed, the deflected dome returns to the ground stable state, and the sensor reverts to its ground state resistance, $R_g$. Consequently, no memory is written into the structure. However, forces at or above this threshold, $F\geq F_{th}$, trigger snap-through to the inverted stable state, and memory of the force application event is encoded in the structure. The strains associated with the inverted state cause the sensor to remain at $R_i$ (Figure \ref{fi:domerestime}). This filtering capability is schematically illustrated in Figure \ref{fi:mechanosensing}d.

To demonstrate tunable thresholding capabilities in our metasheet design, we compare unit cells with domes of two different thicknesses: 0.8 mm and 0.9 mm (Figure \ref{fi:mechanosensing}e). All other parameters are preserved for both domes ($h$ = 7 mm, $r_d$ = 8 mm). Using a tensile testing machine, we apply point loads to the top of each dome (Figure \ref{fi:domeinstron}). When the load is limited to 16 N, $F<F_{th}$ for both domes, so neither dome inverts, and the applied load is filtered out: the sensors return to their original ground state resistance values once the load is removed. When the load is increased to 18 N, the inversion threshold is met for the $t$ = 0.8 mm dome, and the final resistance of the sensor shifts to its inverted state value, thereby demonstrating memory of the event. For the thicker $t$ = 0.9 mm dome, 18 N $<F_{th}$, so the load is again filtered out. By increasing the applied load to 22 N, the inversion threshold is met for the $t$ = 0.9 mm dome, and this is remembered by the change in the sensor resistance. This comparison illustrates the geometry-dependent thresholding that enables filtering of insufficient mechanical signals. The same threshold force tunability can also be achieved by varying the dome height (Figure \ref{fi:domethreshheight}).

\textit{Nonlinear Signal Amplification--} The second key function of mechanosensing units is the amplification of input signals for easier detection. Switching the states of the bistable domes amplifies the local mechanical strain in the metasheet, a nonlinear response that produces large changes in sensor resistance in response to dome inversion.
This amplification is critical for writing memory to the linked memristors; at least a 10\% change in sensor resistance is preferred. To quantify this amplification of local mechanical strain, we conducted finite element analyses (FEA) to examine the stresses and strains in the membrane surrounding the dome, where the sensors are located. These simulations reveal that the stresses and strains in the membrane are inversely proportional to the radial distance from the center of the dome, $r$ (Figure \ref{fi:mechanosensing}b, Figure \ref{fi:domeschematic}b). This relationship between $r$ and the radial and tangential stresses, $\sigma_r$ and $\sigma_{\theta}$, follow the below analytical expressions, where $D_2$, $D_3$, and $D_4$ are constants:~\cite{Udani2021a}

\begin{equation}
  \sigma_r = D_2 (1/r^2) + D_3 + D_4 \log(r)
    \label{eq:sigr}
\end{equation}

\begin{equation}
  \sigma_\theta = -D_2 (1/r^2) + D_3 + D_4 (1 + \log(r))
    \label{eq:sigtheta}
\end{equation}

Therefore, changing the location of the piezoresistive sensor relative to the dome changes the effective amplification factor of the mechanosensing unit, which govens the magnitude of the change in sensor resistance in response to dome inversion. We experimentally demonstrate this first with a basic linear sensor, measuring 24 mm x 1.75 mm x 0.3 mm. Using a dome radius of 8 mm, we print unit cells with sensors placed at 14, 12, and 10 mm from the dome center (Figure \ref{fi:mechanosensing}f). To compare the changes in sensor resistance and remove the variations due to 3D printing, we normalize the resistance values using each sensor’s as-printed resistance. At a distance of 14 mm, there is little discernible change in the sensor’s resistance value when the dome is inverted. At 12 mm, the observed spikes indicate snap-through, but the sensor lacks two distinct resistance levels. At 10 mm, the sensor exhibits both the snap-through spikes and distinguishable ground and inverted state resistance levels. By placing the sensor closer to the dome, we increase the strain in the sensor in the inverted state and therefore its corresponding change in electrical resistance. 

The amplification can be further increased by changing the geometry of the sensor. To target the region of high strain close to the dome perimeter, we print and test sensors with the same dimensions but following circular arcs with varying radii from the center of the dome. At $r$ = 14 mm and $r$ = 12 mm, the spikes indicating snap-through are clearly displayed. Unlike a linear sensor at a distance of 12 mm, the curved sensor displays two distinct levels of resistance, differing in value by 10-15\%. At 10 mm away, the difference between the two resistance levels is 90\% of the initial resistance, compared to only a 40\% difference for the linear sensor. This increase in signal amplification may be explained by inspecting the longitudinal (1-direction) strain in each type of sensor with FEA (Figure \ref{fi:mechanosensing}g). We are most interested in the 1-direction strain because the conductivity is highly directionally dependent due to the internal alignment of conductive microparticles in the direction of extrusion \cite{Santo2021}. The linear sensor shows bands of very low strains, which should not significantly contribute to a change in resistance. Furthermore, the presence of both tensile and compressive strains in the linear sensor causes resistance fluctuations that result in partial signal cancellation. In contrast, the curved sensor shows a uniform compressive strain, with no regions of strain below 1\%. This uniform strain means the effects of dome inversion are more strongly displayed throughout the sensor, resulting in increased signal amplification. For simplification purposes and to avoid the increased damage that occurs with more highly strained curved sensors (Figure \ref{fi:domecrack}), we use metasheets with linear sensors to connect to memristors. Curved sensors could be made more practical by using a less brittle conductive filament; this is left to future work.

\begin{figure}[!t]
  \centering
  \includegraphics[width=\textwidth]{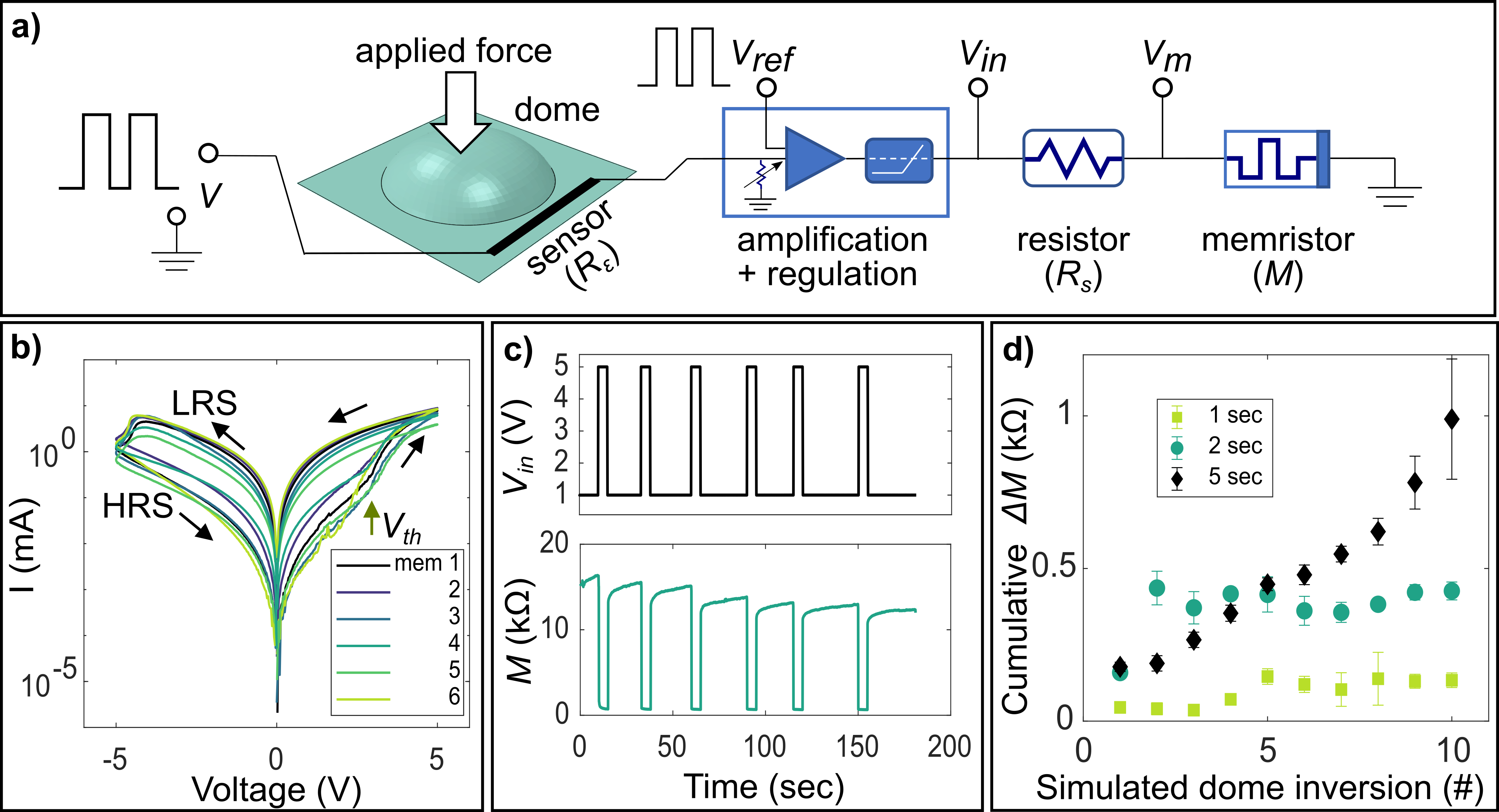}
  \caption{Integration scheme for transducing dome inversions into electrical memories. a) Schematic layout of 1:1 dome-memristor pair connection to incrementally program the memristor in response to changes in dome state. b) Current-voltage relationship for ITO/PEDOT:PSS/Cu memristors shows non-volatile memristance. c) Incremental reductions in memristance (from HRS) induced by 6 subsequent dome inversions corresponding to when the input voltage is 5 V, following the scheme in a). d) Cumulative reduction in memristance with each simulated dome inversion event across different pulse widths.}
  \label{fi:spatial_mem}
\end{figure}

\textit{Spatiotemporal Memory--} The mechanosensing functionality of our metasheets can be used to encode the transduced distributed mechanical inputs into memories. When mechanical inputs cause certain domes to invert, the events' occurrence and spatial pattern are recorded by the metasheet's deflection. The pattern formed in the metasheet indicates the sites experiencing a force exceeding a programmed threshold, $F\geq F_{th}$. 
To catalogue sequentially applied patterns, we couple the dome strain sensors to non-volatile memory resistors, resulting in an auxiliary memory layer. Specifically, this layer consists of voltage-dependent polymeric memristors based on flexible polymer films and off-the-shelf components and aims to electrically remember dome inversion events transduced by the mechanosensing units. Figure \ref{fi:spatial_mem}a illustrates the scheme we used to convert discrete (and even repeated) changes in dome sensor resistance, $R_{\epsilon}$, into multi-bit memristive states, $M$ ($\Omega$). Other integration schemes are also feasible, such as for a 1:1 dome-memristor pair (Figure \ref{fi:spatial_mem}a) or scaled in parallel across larger $m\times n$ arrays. 

Because the dome sensors are passive devices and the memristors are voltage-activated, a fixed amplitude voltage waveform, $V(t)$, is supplied to induce a current through the sensor and a proportional voltage drop across a resistor in series with the sensor. 
A rectangular voltage waveform (where $V_{min}=0$ and $V_{max}>0$) is chosen to reduce the DC power consumption and enable incremental programming of the memristor during times when the dome is inverted (i.e. the applied force is $\geq F_{th}$, and thus the sensor resistance is high: $R_{\epsilon}=R_i$). To enable this signal to program a voltage-controlled memristor, we implemented a fixed gain amplification circuit that outputs a voltage proportional to $\Delta R_{\epsilon}$ and fed this through a switch regulator with a 5 V switching threshold (See Figure \ref{fi:cir_diagram} for details). The switch regulator ensures that the voltage transmitted to the memristor, $V_{in}(t)$, has a consistent amplitude, even when successive dome inversions affect the total change in $R_{\epsilon}$. In this way, $V_{in}(t)$ represents the convolution of $V(t)$ and the time when the dome is inverted (i.e., $R_{\epsilon}=R_i$): the resulting response yields $V_{in}(t)$ equal to 0 V when the dome is not inverted and a peak voltage value of 5 V when the dome is inverted. Voltage signal $V_{in} (t)$ is then applied to the memristor and a resistor wired in series (see SI section \ref{sec:R_sele} for details on how $R_s$ was selected), which enables the divided voltage across the memristor, $V_{m} (t)$, to be sampled as a measurable state of electrical memory for the corresponding dome. Table \ref{tab:logic_scheme} shows the scheme's logic, illustrating that changes in memristance ($|\Delta M|>0$) can only occur when the dome is inverted and $V_m (t)$ (which is also proportional to $V(t)$) is sufficiently high to induce a change in $M$.

\begin{table}[!h]
\centering
\caption{Dome inversion and memristance logic.}
\begin{tabular}{cccccc}\hline
Applied force & Dome state & Sensor resistance, $R_{\epsilon}$ & Input voltage & Memristor  voltage & $|\Delta M|$\\\hline
$<F_{th}$ & Ground  &  Low ($R_g$) & $V_{in}(t)=0$ & $V_{m}(t)=0$ & 0 \\
$\geq F_{th}$ & Inverted  &  High ($R_i$) & $V_{in}(t)\propto V(t)$  &$V_{m}(t)\propto V_{in}(t)$ & $>0$ \\\hline
\end{tabular}
\label{tab:logic_scheme}
\end{table}





The polymeric memristors used herein consist of a PEDOT:PSS thin film ($\sim$ 60 nm thick) sandwiched between indium tin oxide (ITO) and copper (Cu) electrodes (effective electrode diameter $\sim$ 1.8 mm; see Figure \ref{fi:polymeric_film} for device architecture). PEDOT:PSS forms a strong, flexible polymer matrix~\cite{Vos2012,Fan2016} that enables mixed ionic-electronic transport, and it has been used to create organic memristive devices interfaced with reactive and noble metal electrodes displaying high OFF/ON ratios~\cite{Moller2003,Ha2008}. Furthermore, the fact that PEDOT:PSS films can be fabricated through simple, low-temperature processing to make functional memristors makes it a promising memristor material candidate for enabling a conformal memory layer in mechanosensitive neuromorphic metasheets.

Representative voltage-controlled current measurements obtained on six distinct PEDOT:PSS memristors are shown in Figure \ref{fi:spatial_mem}b. The arrows denote the direction of the measured current path in response to one cycle of a 100 mHz sinusoidal voltage. The significant hysteresis in these $i-v$ curves indicates that the devices can reversibly switch between a high resistive state (HRS $\sim$ 15.2 k$\Omega$) and a low resistance state (LRS $\sim$ 1.31 k$\Omega$) in response to varying the applied voltage. The counterclockwise direction of current flow at positive voltages and the clockwise direction of current flow at negative voltages shows that these memristors exhibit a non-volatile memristance.~\cite{Hopkin2008,Chua19}  Specifically, when the applied voltage on the Cu electrode relative to the grounded ITO electrode increases above +3 V ($V_{th}$) (See \ref{fi:HRS_LRS} Figure for measurement technique details), the devices switch (i.e., SET) from the HRS to the LRS as reflected in the sharp increase in current. Only when the probe voltage is lowered below -4 V do the devices RESET to the HRS. This means that programmed states of resistance can be stored in the sample even when the voltage to the device is removed  (or set to zero). While the precise mechanism of voltage-activated resistive switching in these devices is still under investigation, these simple, low-cost, potentially-conformal devices yield consistent performance across dozens of batches (Figure \ref{fi:HRS_LRS}) and suitable cycle-to-cycle stability (Figure \ref{fi:Mem_cycles}).  

The 1:1 dome-memristor pair operation scheme described above and shown in Figure \ref{fi:spatial_mem}a seeks to incrementally reduce from HRS the memristance of a specific memory device in response to discrete dome inversions. Based on the $i-v$ behaviors in Figure \ref{fi:spatial_mem}, we tested the ability to incrementally program the memristors using a series of 5 V  rectangular pulses. Each pulse lasts 5 sec and occurs once every 15 sec, an intentional choice that corresponds closely to the 100 mHz sweep frequency used for $i-v$ measurements (Figure \ref{fi:diff_frequency}). A low-amplitude voltage ($1$ V $< V_{th}$) is used to \textit{read} the new resistance state of the memristor between pulses. Figure \ref{fi:spatial_mem}c shows the incremental reductions in $M(t)$ induced by successive 5 V pulses, i.e., it is partially switched from HRS to LRS with each pulse. 
Adjusting the pulse width and tuning the pulse amplitude relative to the SET voltage can be used to design the memristance variation per event and extend the number of programmable resistance states that each memristor can access. By varying the amplitude of $V_{in} (t)$ from the switch regulator, we observed that a 4 V amplitude of $V_m$ results in a steady cumulative change in resistance across subsequent pulses (Figure \ref{fi:cum_change}). For $V_m$ values below 3.5 V, we observe smaller total changes and greater variability per subsequent pulse. This information allowed us to determine the necessary amplitude of $V_{in}(t)$ supplied by the switch regulator. We also explored the duration of the writing pulses to determine a suitable pulse width. We compared the cumulative reductions in a memristor's resistance for three different pulse widths ($T_{on}$), revealing that for 1 and 2 second pulses, the cumulative changes in the resistance are  small and unsteady (Figure  \ref{fi:spatial_mem}d). This is because the pulse is too brief to drive adequate memristance changes. In contrast, using 5 sec pulses yielded consistent, monotonic changes in device resistance, validating the 1:1 dome-memristor operational scheme. While this writing pulse is long compared to other memristors, we expect they could respond significantly faster by reducing the thickness and electrode dimensions, and unveiling the mechanisms of switching. For  comparison, PEDOT:PSS-based organic electrochemical transistors can exhibit voltage-driven conductance changes in less than one millisecond.~\cite{Khod2011}


\textit{Spatially distributed input learning: physical Hopfield networks--} 
The combined mechanosensing and history recording capabilities of our metamaterial allow us to implement a physical learning system based on Hopfield networks~\cite{Hopfield1982} and the Hebbian learning rule.~\cite{Rojas1996,Hertz2018} Hopfield networks consist of interconnected neurons and synapses that learn a series of input patterns stored as minima (i.e., attractors) in their energy landscapes. These patterns can be retrived following the associative memory paradigm.~\cite{Rojas1996} Hopfield networks achieve this by utilizing a fully connected network architecture that strengthens the  synaptic connection between activated artificial neurons. Specifically, patterns ascribed to neuron states, such as images represented by pixels, are stored in the network's interaction matrix ($\mathbf{J}$) (SI section \ref{sec:Asso_Memory} for details), which form different attractors in the networks’ energy landscape (Figure \ref{fi:Asso_mem}).~\cite{Hertz2018}  We leverage the cumulative resistance changes in our memristors, triggered by dome inversions, to capture the interactions between dome units, thus updating the synaptic weights $J_{ij}$ and storing the spatial patterns sequentially input into our metamaterial. 

\begin{figure}[!h]
  \centering
  \includegraphics[width=\textwidth]{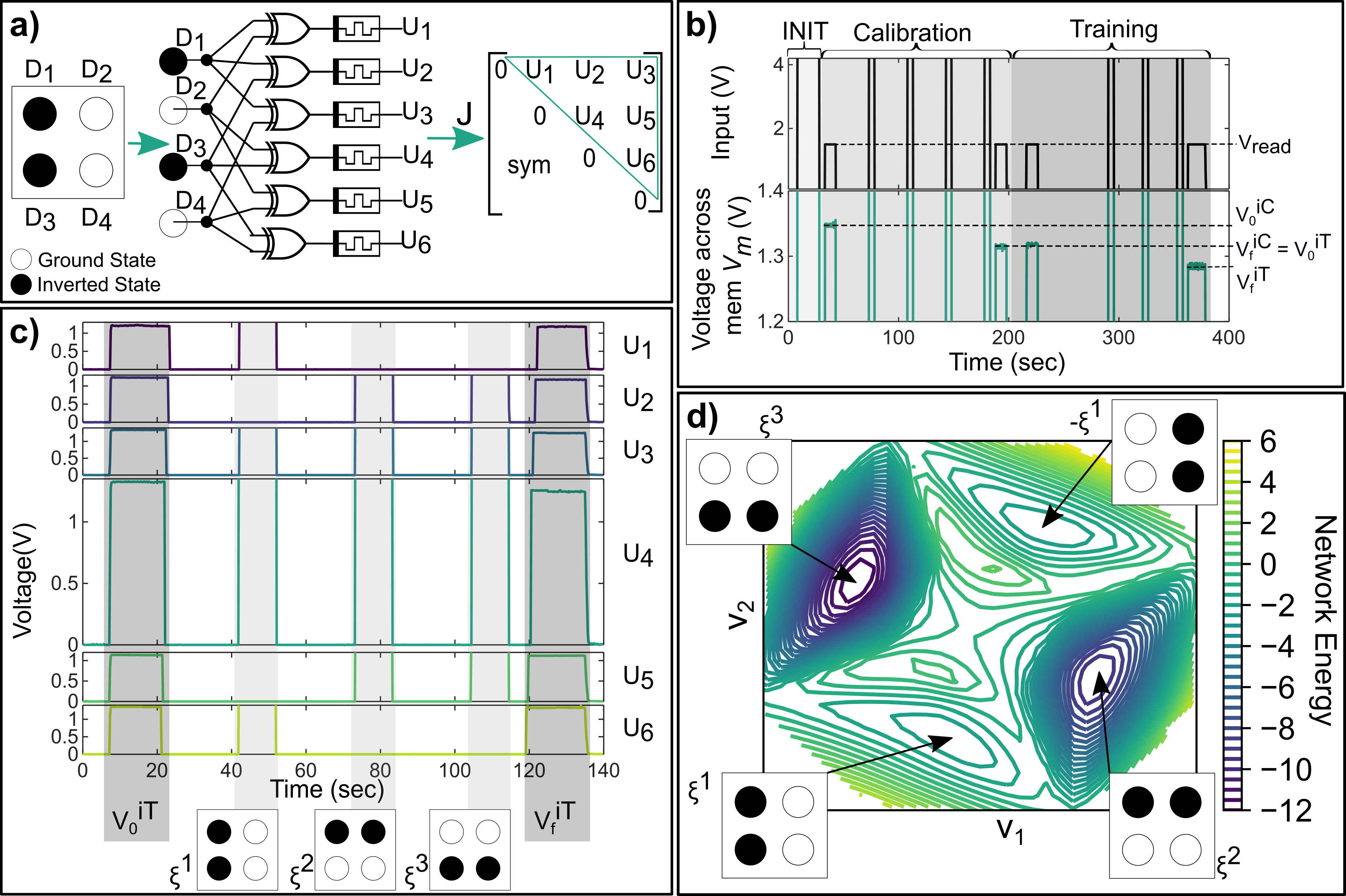}
  \caption{Physical Hopfield network implementation and test for different input patterns. a) Schematic of physical Hopfield network and connection between memristors to capture the interactions between mechanosensor units. b) 3-stage scheme designed to capture the interaction between mechanosensors with memristor response. c) Voltage across six different memristors, input patterns (light gray), and initial and final voltage reads (dark gray). The $y-$axis of $U_4$ is extended to show the difference between voltage reads. d) The Hopfield network energy landscape projected into the $(V_1, V_2)$ plane by dimensionality reduction shows the input patterns as energy minima/attractors. Input patterns  $\in R^4$ are visualized using the directions with the two largest variances ($V_1, V_2$ $\in R^2$).}
  \label{fi:hopfield_net}
\end{figure}

The physical Hopfield network consists of a $m \times n$ dome metasheet array (Figure \ref{fi:hopfield_net}a), with each unit cell (dome + sensor) acting as an individual artificial neuron\footnote{The term artificial neuron refers to a unit in Hopfield networks jargon; however the terminology does not imply biomimetic neural functionality.} and the interconnecting memristors acting as synapses. Each unit can adopt two possible states: (+1) ground state and (-1) inverted state (Figure \ref{fi:mechanosensing}a). The interactions between unit cells can be captured by connecting different dome pairs to an XOR gate, as illustrated for a $2 \times 2$ ($n=m=2$) array in Figure \ref{fi:hopfield_net}a. This writes a resistance change for every “on” and “off” neuron pair interaction (Figure \ref{fi:Mem_interac_measurment}). The number of memristors was reduced by considering the convergence properties of the networks ($J_{ii}\leq 0$ and $J_{ij}=J_{ji}$), which reduces the number of needed interactions to the upper triangular positions of the $mn \times mn$ interaction matrix labeled as $U_i$ ($i = 1,2,\hdots, M=\frac{ m\times n }{2}(m\times n - 1)$; see Figure \ref{fi:hopfield_net}a). By using this scheme, the Hopfield network can be trained directly by physical inputs with different external patterns as they occur (i.e., the network is trained online).

A 3-stage measurement scheme was implemented to capture the interactions between dome units by using the cumulative memristor response (Figure \ref{fi:hopfield_net}b). In stage 1, an initialization pulse (INIT) of 4.3 V for 20 sec was used to pre-set each memristor to a value that changes linearly in response to subsequent voltage pulses (described in Figure \ref{fi:Cu/PSS_Response}). Stage 2 involves a calibration procedure consisting of a series of four 4.3 V, 5 sec writing pulses (simulating dome inversion events) preceded and followed by 1.5 V, 10 sec reading pulses. The voltage across the memristor ($V_m$) is measured correspondingly, which directly reflects the resistance state of the memristor. The first read pulse reflects the initial HRS of the memristor ($V^{iC}_0$), and the second read pulse reflects the resistance state of the memristor after the four calibration writing pulses ($V^{iC}_f$). The difference between the two read pulses accounts for the net change $\Delta V^i = (V^{iC}_0-V^{iC}_f)/4$ due to four simulated dome inversions. In stage 3, the memristor network is subjected to a series of training pulses generated by actual dome inversion events (physically induced), which are also preceded and followed by reading pulses. The measured difference in $V_m$ read between the third ($V^{iT}_0$) and final ($V^{iT}_f$) read events reflects the net change in $V_m^T$ (resistance) due to pair-wise interactions between dome inversion events. The entries to the interaction matrix are calculated as $U_{i} = \text{round}\left(\frac{V^{iT}_0-V^{iT}_f}{\Delta V^i }\right)$. By performing this procedure across the six memristors in parallel, we can automatically store and update interaction values between units (Figure \ref{fi:hopfield_net}c), generate the interaction matrix between dome units, and learn the input patterns using the physical Hopfield network.

We demonstrate the Hopfield network's online learning capability with a $2 \times 2$ metasheet array ($n=m=2$, with unit cell dimensions $w$ = 30 mm, $h$ = 7 mm, $t$ = 0.8 mm, $r_d$ = 8 mm; see Figure \ref{fi:hopfield_net}). The array was trained by applying a sequence of patterns using physical dome inversions (see Movie S1). The resulting inter-unit firing is captured by constructing the interaction matrix from the memristor voltages read before and after the patterns are input (Figure \ref{fi:hopfield_net}c dark gray shaded region). As the Hopfield network's interaction matrix stores several patterns additively in the final values of the weights (see SI section \ref{sec:Asso_Memory}), only the differences between initial and final memristance measurements are necessary to retrieve the stored collection weights of the trained network. Once the interactions are stored, the network's weights are set, and the unique patterns are retrieved offline by presenting corrupted images to the network and performing an energy minimization process with asynchronous neuron update (see SI section \ref{sec:Asso_Memory}).

The training performance was evaluated by examining the Hopfield network's energy landscape to determine whether the physically input patterns were successfully stored as energy minima (Figure \ref{fi:hopfield_net}d). To achieve this, we utilized a dimensionality reduction technique~\cite{Kang2021}  that uses random data generated based on the trained interaction matrix and the energy function to visualize the network's attractors using the two largest variance directions ($V_1$ and $V_2$). This procedure reveals four different attractors within the landscape (Figure \ref{fi:hopfield_net}d), each corresponding to the input patterns ($\xi^i$) and their reflections ($-\xi^i$). This behavior is expected as the Hopfield network's energy function is quadratic, yielding equal values for $\xi^i$ and $-\xi^i$. This leads to the system learning both configurations (i.e., $\xi^i$ and $-\xi^i$) since both are energy minima. Moreover, it is worth mentioning that $\xi^2$ and $\xi^3$ are the reflections of one another. This implies that during the training process the network learns this pattern twice, resulting in deeper energy wells (see Figure \ref{fi:hopfield_net}d). The obtained results indicate that the training procedure captured all interactions between dome units, and the metasheet's memory layer successfully remembers the input patterns. Classification accuracy of the physical training is evaluated by presenting 3000 different corrupted patterns (Figure \ref{fi:Asso_mem}) to the network and then determining the number of correct identifications (i.e., implying no errors with respect to the stored patterns during training). A 91$\%$ overall accuracy is found with the pattern sequence shown in Figure \ref{fi:hopfield_net}c (see SI section \ref{sec:Hopfield_stat}), which has a  5\% difference from the offline trained Hopfield network (Table \ref{tab:net_accuracy}). These results show that one of the learned patterns during the physical training is retrieved with no error 91$\%$ of the time a pattern is tested. The energy minimization process entailing the weights' updating when several corrupted patterns are presented to our physically trained network is shown in Movie S2.


\textit{Conclusions--} We introduce a new class of flexible metamaterials which demonstrate spatiotemporal mechanosensing and neuromorphic functionalities. Our prototype leverages the structural bistability of dome-shaped units to filter, amplify, and transduce external mechanical signals into electrical states that are used to induce non-volatile (permanent) changes in memristance. The proposed architecture allows for producing neuromorphic properties in metamaterials that can cover large areas (10$^2$ cm$^2$), in contrast to many examples of neuron-inspired sensing and perceptual systems. We use the spatial organization of our dome units and the cumulative changes in memristance of our metamaterial to realize a physical Hopfield network capable of learning a series of mechanical input patterns online. Specifically, we leverage the bistability-based filtering and amplification of our dome units to generate threshold-dependent changes in sensor resistance that can themselves be used to program a corresponding memristor state. We use the memristance value to populate associated interaction matrix weights, thus capturing the firing of units following a well-known Hebbian rule. Crucially, the cumulative resistance changes of our metamaterial’s memristors allow for retrieving all the stored patterns from their final states. As a result, our metamaterial encodes several input patterns without the need to store temporal information. In this sense, our metamaterial exhibits analogous behavior to threshold-dependent event cameras augmented with a type of memory that encodes the captured episodes directly into the physical network's weights.  We envision this unique capability as providing a route to continuously learning new spatiotemporal mechanical inputs which can be compared to known patterns of interest that may trigger specific control actions. For example, known pressure patterns associated with dangerous aerodynamic conditions could be presented to the current state of the physical Hopfield network connectivity matrix (see equation S.6). Reaching an energy minimum would indicate that such a pattern has been learned (see Figure \ref{fi:conceptual}), indicating with minimal computational cost that a control action is needed. Thus, our mechanosensitive neuromorphic metamaterials reduce the need for costly online data storage, transmission, and processing of spatiotemporal inputs in situations where tactile pattern identification is of interest, including for robotics, autonomous systems, wearables, and morphing wings.

\section*{Materials and Methods}
\textit{Unit cell and metasheet fabrication--} The unit cells are 3D printed using fused deposition modeling (FDM) on a Raise3D Pro2 printer using the settings in Table \ref{t:domesprint}. We use Ninjatek Cheetah TPU filament (E = 26 MPa) \cite{NinjaTek} for the dome and base and BlackMagic3D PLA + graphene composite filament (E = 3767 MPa) for the sensor \cite{Graphene3DLab2015,Mansour2019}. Due to the PLA + graphene material being two orders of magnitude stiffer than the TPU, the sensors are embedded in the TPU base to avoid delamination. A connection point is left exposed at each end. To address the high contact resistance \cite{Cao2020,Schouten2020}, we paint each connection point with a thin layer of highly conductive carbon paint \cite{Pelco} before attaching copper wire, using another layer of conductive paint. After drying, we apply a dot of hot glue on each connection to secure the wire in place. Metasheets are manufactured following the same procedure, with one continuous 3D print.

\textit{Unit cell resistance measurements--} The resistances of the dome sensors are measured using a voltage divider circuit with a 2 k$\Omega$ shunt resistor. The voltage drop over the sensor is measured using a National Instruments USB-6251 data acquisition system and a simple National Instruments Labview multimeter program. The DC power supply (Keysight E36313A) applies 2 V to the circuit.

\textit{Mechanical tests--} Inversion force and thresholding tests are performed on an Instron 3345 Universal Testing Machine with Bluehill software. Test domes are printed with a 2 mm diameter hole in the center, where an M2 screw is attached with a nut and washer. The other end of the screw attaches to a machined aluminum block, which is held by the Instron machine grips. The dome sample to be tested sits on a 3D printed base with a 20 mm x 20 mm square hole centered under the dome to allow for free inversion (Figure \ref{fi:domeinstron}). The sensor is hooked up to the same voltage divider circuit as used for other resistance measurements. Resistance data is gathered for 10 seconds before the Instron test begins in order to capture the initial state of the sensor. For each test, the head of the machine moves down at a rate of 20 mm/min. Once inversion is detected via a drop in the reaction force, the test stops, and the machine returns the dome to the starting position. The sensor data is collected until the total test time reaches 2 minutes. The dome is then reset manually and the test rerun. Each specimen tested first undergoes 10 full inversion tests. Then, the threshold behavior is tested following the same settings as for the inversion force, but with the end of test triggered by a maximum applied load. This maximum load is incrementally increased each test until the inversion force threshold is reached. 

\textit{Finite element analysis--} The unit cells are modeled in Abaqus using linear elastic material properties and S4R shell elements. To capture the bistable behavior, geometric nonlinear analysis is used. The unit cell is initially modeled in the stress-free ground state. Snap-through is triggered using an enforced displacement applied to the center node of the dome while the edges are pinned. These boundary conditions are then released while the center node is fixed to show the free inverted state of the unit cell.

\textit{Fabrication of memristors--} ITO glass slides (10-15 $\Omega$/sq) were cleaned with (in the following order) soap-water, de-ionized water, ethanol, acetone, propanol via sonification for 15 minutes each and dried using ultra-pure Nitrogen. The conductive face of clean ITO slide was oxidized in an oxygen plasma chamber using a PE 50 XL Benchtop low-pressure plasma system for 1 minute to increase the hydrophilicity of the surface, which encourages wetting of PEDOT:PSS solution resulting in uniform distribution. 3 wt $\%$ PEDOT:PSS solution bought from Sigma-Aldrich was spin-coated on to a clean ITO glass slide at 3000 rpm for 40 seconds. It was then dried at 120 $^{\circ}$C for 20 minutes and vacuum annealed for more than 24 hours. This results in a thin film ($\sim$ 60 nm thickness) measured using a F20 device from Filmetrics. Portions of the ITO were covered with insulating tape, which when removed post-fabrication recovered a naked ITO surface to be used as a bottom electrode as seen in Figure  \ref{fi:Resistance_C}b.

\textit{Characterization of memristors--} Electrical characterization of the fabricated devices was done using a SP 200 Potentiostat from Biologic. Spring-loaded copper contacts were used as the top electrode as shown in Figure \ref{fi:Resistance_C}b. Bipolar sinusoidal voltage waveforms of 5 V and 100 mHz were used for the current-voltage (i-v) characterization of the devices.

\textit{Programming of pulsatile input--} A function generator (Hewlett–Packard 3314A) was used to generate unipolar square voltage pulses as supply V(t) and $V_{ref}$ to the circuit. Multiple Op-Amps were used to build the amplifier circuit as shown in Figure S10. A DC power source (BK precision Triple Output DC power source 1762) was used to tune $V_{in}(t)$ for reading and writing pulses. An Arduino (DUE R3) was used to record the response signals across the memristor, i.e $V_{m}$.

\section*{Author Contributions}
All authors designed the research. KSR, HM, and JPU developed the initial sensor design. KSR developed the metasheet manufacturing method and conducted mechanosensor unit characterization. KSR and JCO conducted FE simulations of the mechanosensors. SK, YY, and SAS conducted memristor material development, characterization, and electrical activation. JCO, KSR, and AFA developed the Hopfield network concept and JCO conducted the simulations. SK, JCO, and KSR developed the physical implementation of the Hopfield network, and SK conducted the experiments. All authors contributed to the writing and revision of the manuscript and supplementary materials at different stages, and AFA and SAS provided guidance throughout the research.

\bibliographystyle{is-unsrt}
\bibliography{PostDocDatabase,references}

\appendix

\renewcommand{\theequation}{S.\arabic{equation}}
\renewcommand{\thefigure}{S\arabic{figure}}
\renewcommand{\thetable}{S\arabic{table}}
\setcounter{figure}{0}
\setcounter{table}{0}

\section{Supplementary Material}\label{supplementary}

\subsection{Mechanosensing Unit Cell}

\begin{figure}[!h]
  \centering
  \includegraphics[width=0.9\textwidth]{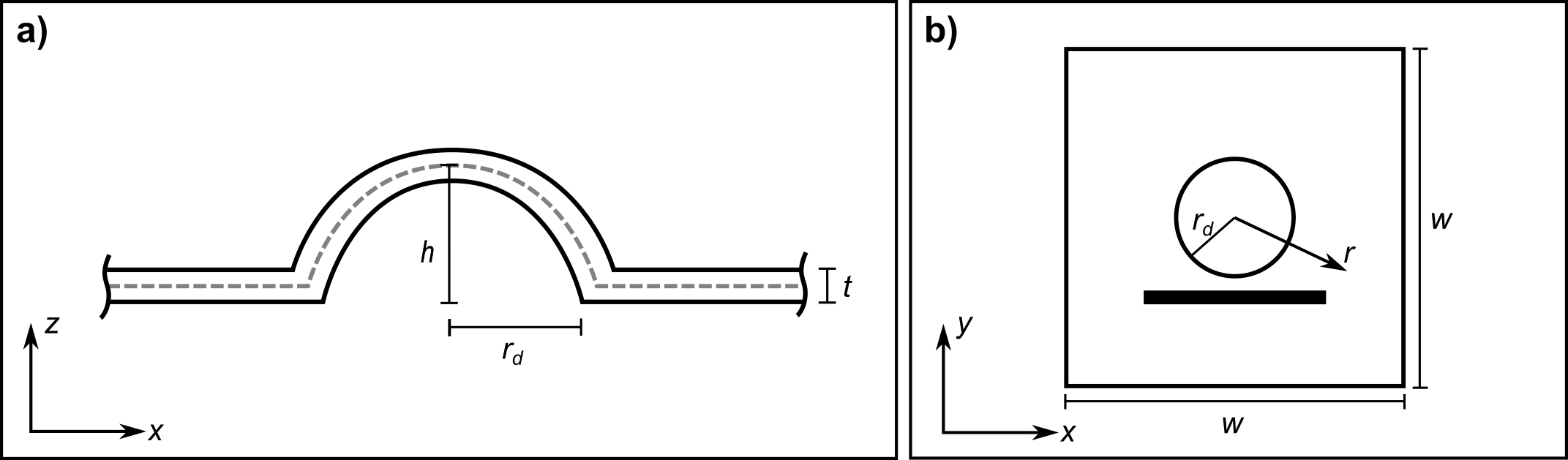}
  \caption{a) Cross-section of the dome showing the height, $h$; thickness, $t$; and radius, $r_d$ b) The dome is in the center of a $w\times w$ unit cell; $r$ is the distance from the center of the dome.}
  \label{fi:domeschematic}
\end{figure}

The key dimensions of the dome unit cell are shown in Figure \ref{fi:domeschematic}. Print settings are listed in Table \ref{t:domesprint}.

\begin{table}[h]
\centering
\caption{Mechanosensor print settings}
\resizebox{.25\textwidth}{!}{%
\begin{tabular}{|ll|}
\hline
\multicolumn{2}{|l|}{TPU}                            \\ \hline
\multicolumn{1}{|l|}{Nozzle Diameter}      & 0.40 mm \\ \hline
\multicolumn{1}{|l|}{Layer Height}         & 0.10 mm \\ \hline
\multicolumn{1}{|l|}{Extruder Temperature} & $230 ^{\circ}C$   \\ \hline
\multicolumn{2}{|l|}{PLA + graphene}                 \\ \hline
\multicolumn{1}{|l|}{Nozzle Diameter}      & 0.40 mm \\ \hline
\multicolumn{1}{|l|}{Layer Height}         & 0.10 mm \\ \hline
\multicolumn{1}{|l|}{Extruder Temperature} & $200 ^{\circ}C$   \\ \hline
\multicolumn{2}{|l|}{Print Bed}                      \\ \hline
\multicolumn{1}{|l|}{Bed Temperature}      & $50 ^{\circ}C$    \\ \hline
\multicolumn{1}{|l|}{Bed Material}         & Glass   \\ \hline
\end{tabular}
}
\label{t:domesprint}
\end{table}
\newpage

\subsection{Dome Behavior}
\begin{figure}[!h]
  \centering
  \includegraphics[width=\textwidth]{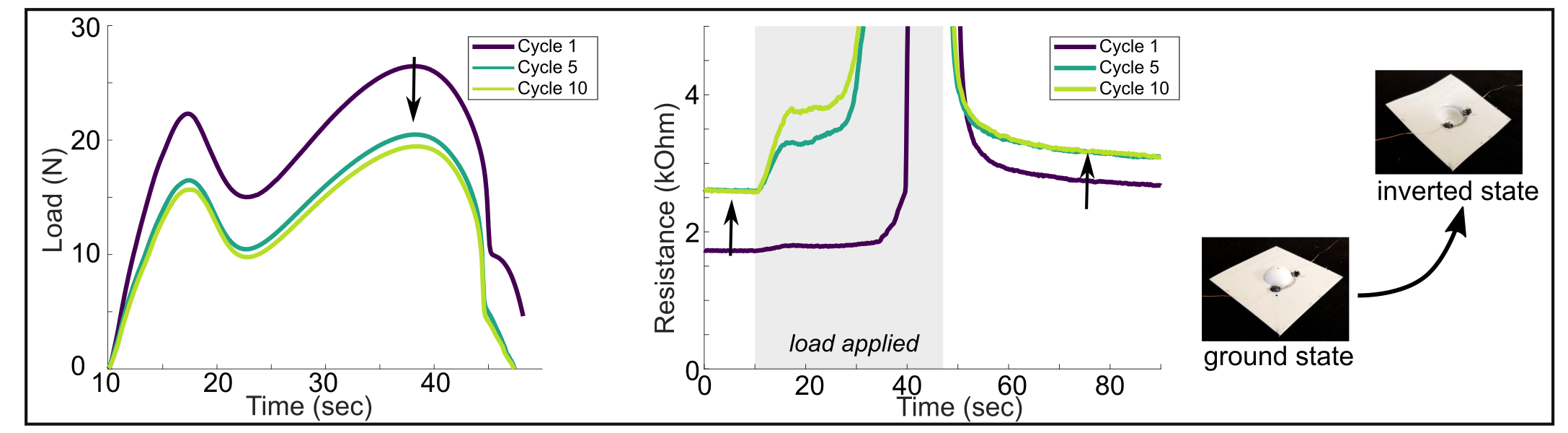}
  \caption{Typical of viscoelastic structures, the first loading cycle is always noticeably different than subsequent cycles. This stress-softening is reflected in the resistance values of the sensor.}
  \label{fi:domecycle1to10}
\end{figure}

\begin{figure}[!h]
  \centering
  \includegraphics[width=\textwidth]{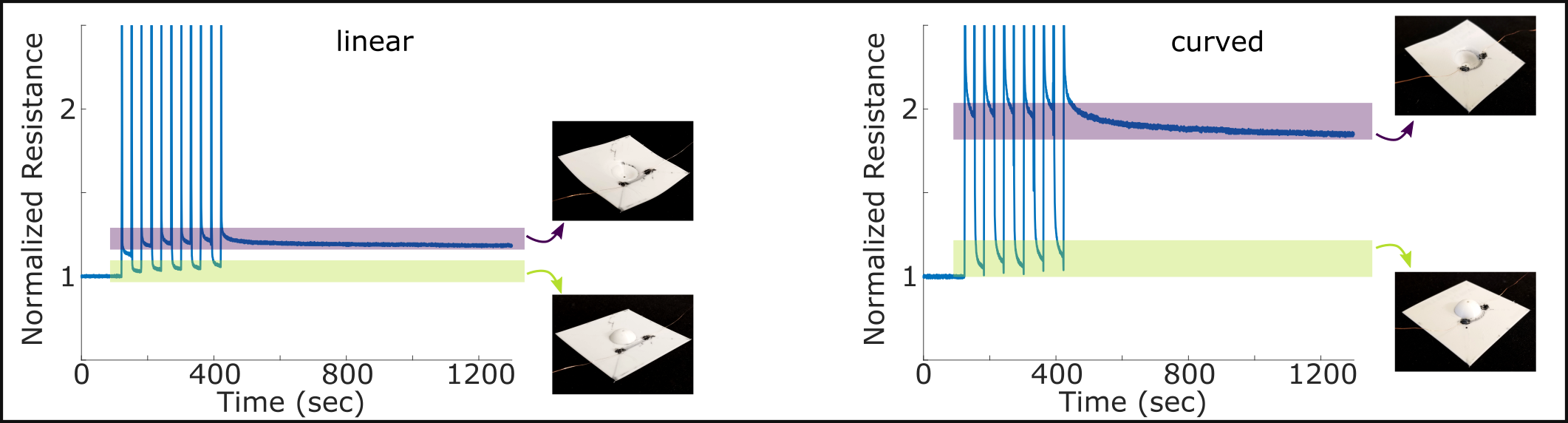}
  \caption{The resistance of the inverted state is maintained over time.}
  \label{fi:domerestime}
\end{figure}

As shown in Figure \ref{fi:mechanosensing}c, the first loading cycle generally shows different behavior from subsequent cylces for a mechanosensing unit cell. This is due to the stress-softening behavior, as shown in Figure \ref{fi:domecycle1to10}.\cite{Christ2017} The ground and inverted state resistances are steady over time, showing the mechanosensor retains information about whether or not an inversion has occurred (Figure \ref{fi:domerestime}). 

\subsection{Instron Tests}
\begin{figure}[!h] 
  \centering
  \includegraphics[width=.3\textwidth]{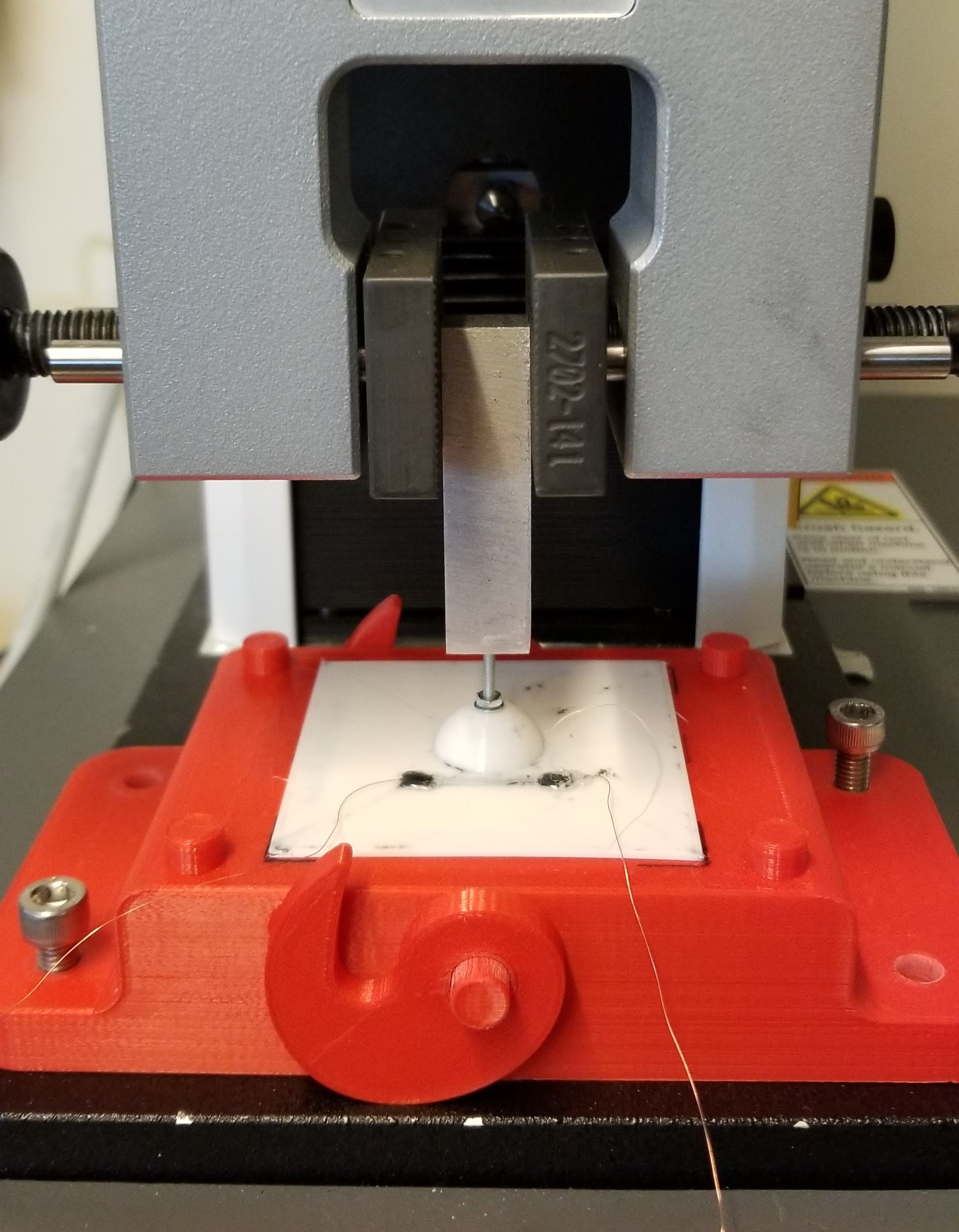}
  \caption{Instron testing setup for measuring snap-through load of domes.}
  \label{fi:domeinstron}
\end{figure}

\begin{figure}[!h]
  \centering
  \includegraphics[width=\textwidth]{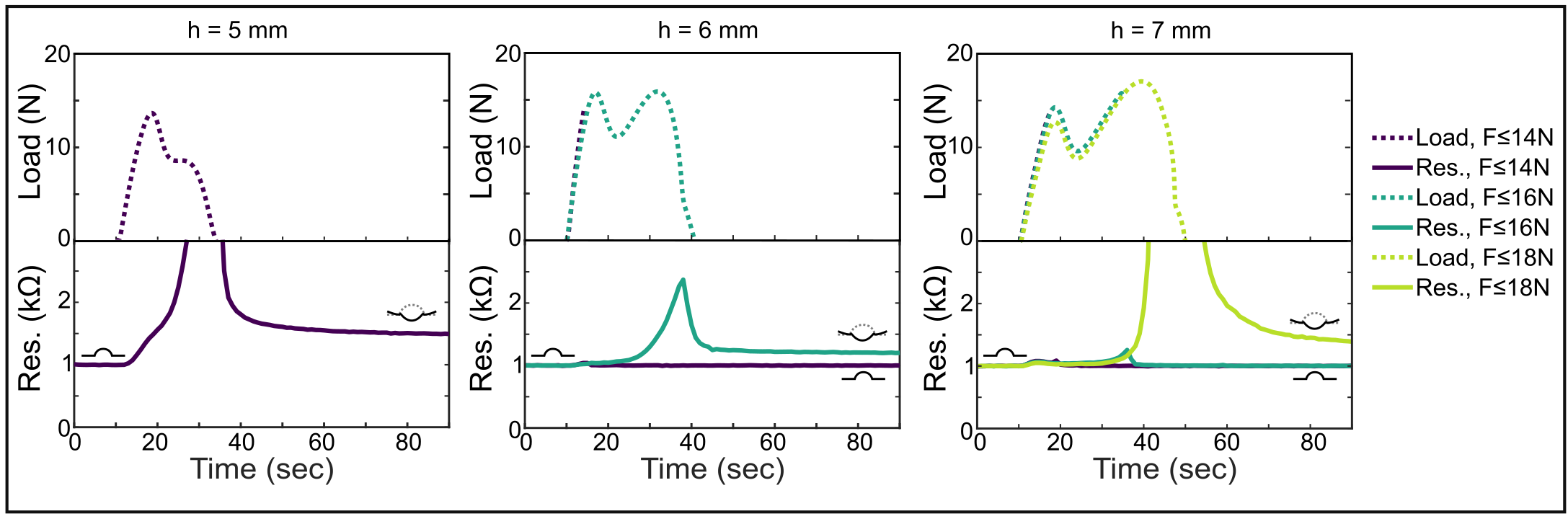}
  \caption{The filtering tunability is demonstrated by the changing snap-through load for three different heights for domes of thickness 0.8 mm. Loads below this threshold load are not recorded by the sensor.}
 \label{fi:domethreshheight}
\end{figure}

The Instron test setup used to demonstrate the thresholding capability is shown in \ref{fi:domeinstron}. The threshold load for filtering may be determined by changing the geometric parameters of thickness (Figure \ref{fi:mechanosensing}d,e) and height (Figure \ref{fi:domethreshheight}).


\subsection{Curved Sensors}
\begin{figure}[!h]
  \centering
  \includegraphics[width=.75\textwidth]{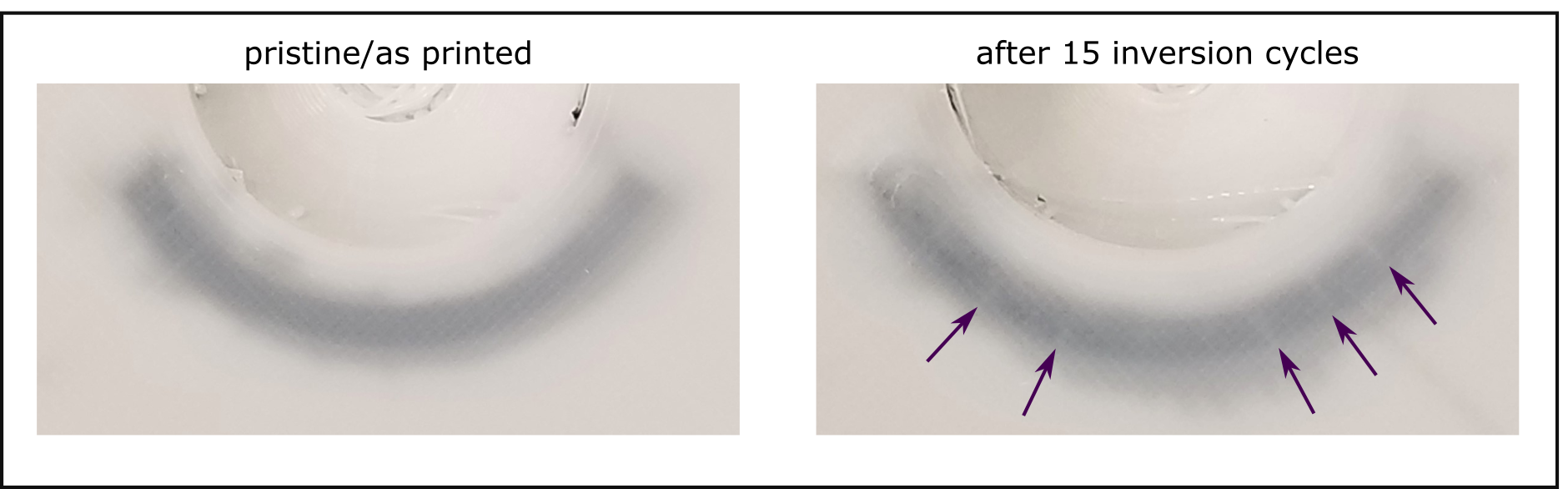}
  \caption{Curved sensors are more prone to cracking after repeated inversion cycles due to the higher strains. A pristine/as printed sensor (left) is compared with a sensor after 15 inversion cycles (right). Arrows indicate visible crack locations}
  \label{fi:domecrack}
\end{figure}

While curved sensors offer significantly higher signal amplification, there are several practical drawbacks. Because they follow a region of higher stress and strain, they are more prone to cracking (Figure \ref{fi:domecrack}) and failures at the connection points. While the damaged sensors may still function, the damage alters the ground state resistance. The curved shape also makes them more sensitive to the hysteresis and viscoelastic effects of the TPU base; this can be seen in changing slopes of the resistance after snap-through and the gradual increase in the ground state resistance over cycling. Furthermore, the conductivity of printed conductive filament is highly dependent on the internal alignment of the conductive microparticles \cite{Santo2021}. The effects of printing along a curve rather than in a straight line on this alignment are not fully understood. 

\newpage


\subsection{Circuit diagram for signal amplification}\label{sec:Amp_Circuit}

The model of Op-Amp is LM741, which is the most used model. OpAmp1 is used as a voltage follower circuit, which can also be considered as a buffer amplifier. Here, the op-amp does not amplify the input signal, and the gain is 1. Therefore, the output voltage is same as the input voltage. Because the voltage follower circuit has a very high input impedance, it can provide a stable voltage output, which follows the input voltage. In this circuit design, the input voltage is decided by $V_{in1}$ and the resistance ratio between $R_{ref}$ and resistance of the dome ($R_{\epsilon}$). Originally, the $R_{ref}$ is adjusted to equal to dome resistance. Therefore, the output voltage is half the input voltage ($V_{in11}$ = $0.5V_{in1}$).  Whereas OpAmp2 is a typical differential amplifier circuit. The $V_{out}$ is defined as:

\begin{equation}\label{eq:Vout1}
  V_{out} = \left(\frac{R_1 + R_4}{R_5 + R_6}\right)\left(\frac{R_4}{R_1}\right)V_{in2} - \frac{R_4}{R_3}V_{in11}
\end{equation}

When $R_1 = R_3= 100 {\Omega}$ and $R_2 = R_4= 2000 {\Omega}$. The equation can be simplified as:

\begin{equation}\label{eq:Vout2}
  V_{out} = \frac{R_2}{R_1}\left(V_{in11}-V_{in2}\right)
\end{equation}

From the equation, it can be seen that the OpAmp2 circuits are amplifying the difference between $V_{in11}$ and $V_{in2}$. In our design, $V_{in2}$ is set to 1 V and $V_{in11}$ is $\frac{1}{2}V_{in1}$ , which is 1 V as well. Therefore, originally, there is no voltage output from op-amp circuits. By inverting the dome, the resistance of the dome changes, which leads to a change in resistance ratio between $R_{\epsilon}$ and $R_{ref}$. This event will change the $V_{in11}$ because of the characteristic’s voltage follower circuits. Therefore, a voltage difference will be created between $V_{in11}$ and $V_{in2}$. This voltage difference will be picked up and amplified by the OpAmp2 circuits.

\begin{figure}[!h]
  \centering
  \includegraphics[width=0.65\textwidth]{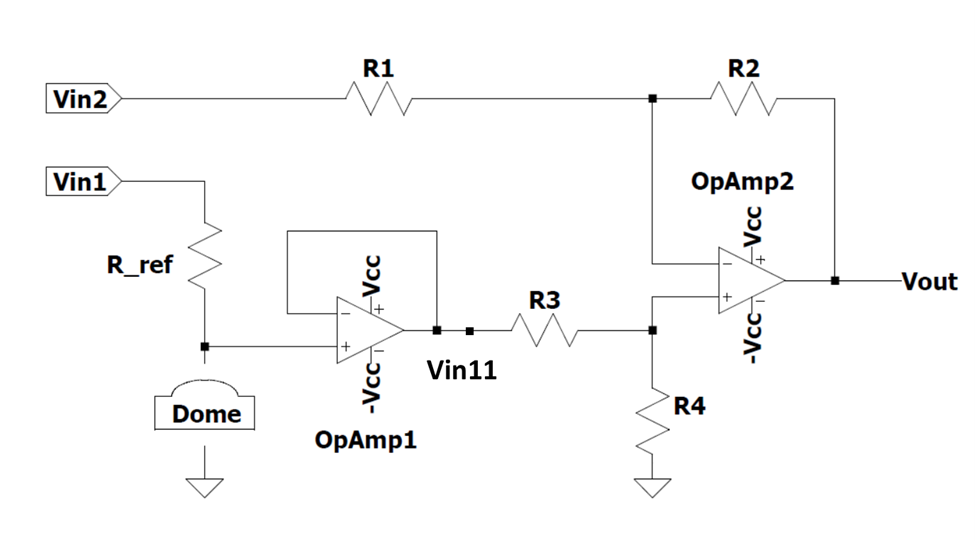}
  \caption{Circuit diagram for amplification of input signal.}
  \label{fi:cir_diagram}
\end{figure}

\subsection{Scheme for selection of resistor in series with memristor}\label{sec:R_sele}

\begin{figure}[!h]
  \centering
  \includegraphics[width=0.65\textwidth]{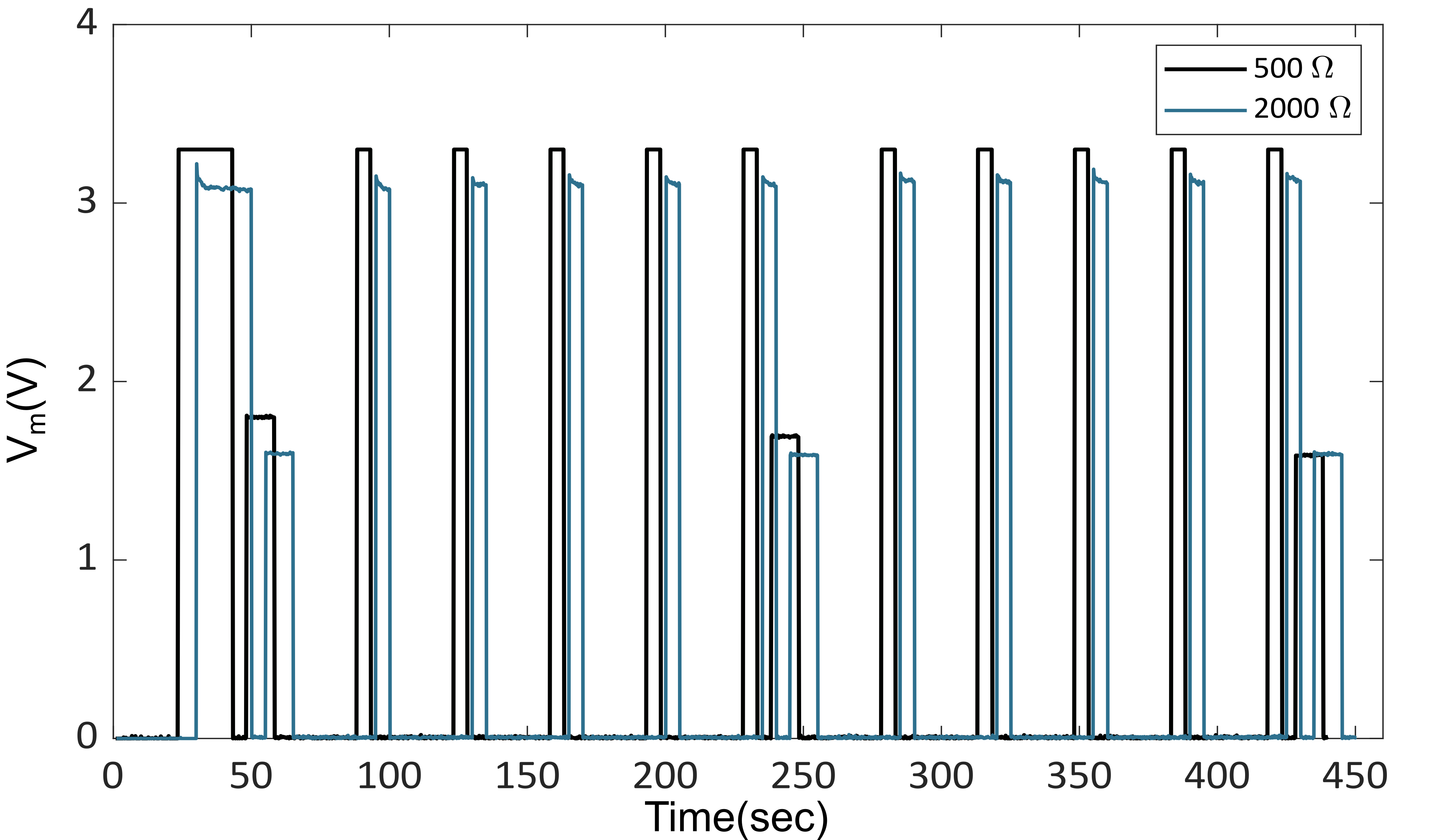}
  \caption{Selection of a resistor in series to achieve distinguishable change in $V_m$ ($\sim$ resistance of memristor).}
  \label{fi:Resistance_C}
\end{figure}

We use a resistor in series with the memristor to read the voltage across the memristor $V_m$ by using an Arduino circuit. Therefore, a careful choice of a resistor in series is required for ensuring i) majority of the voltage amplitude ($> V_{th}$) from the pulse is felt across the memristor for effective drop in resistance, ii) resolution of change in $V_m$ is adequately above any noise level. A 500 $\Omega$ and 2000 $\Omega$ resistor was tested in series with the memristors. Initially, a pulse of 4.3 V was applied for 20 seconds to the resistor-memristor duo to drive the first big non-linear change in the memristor and set the memristor in the more linearly increasing conductive region. This was followed by cycles of a read pulse (1.5 V, 10 seconds) and five write pulses (4.3V, 5 seconds). Figure \ref{fi:Resistance_C} compares the voltage drops across the memristor ($V_m$) using the read pulses, which are proportional to the current resistive state of the memristors. It’s observed when the 500 $\Omega$ resistor is connected in series, upon subsequent voltage pulses, there is a significant drop in read voltage compared to the 2000 $\Omega$. These distinguishable drop in $V_m $ above noise level represent change in memristance ($M$).

\subsection{Device architecture and electrical characterization}\label{sec:film_arc}
\begin{figure}[!h]
  \centering
  \includegraphics[width=0.9\textwidth]{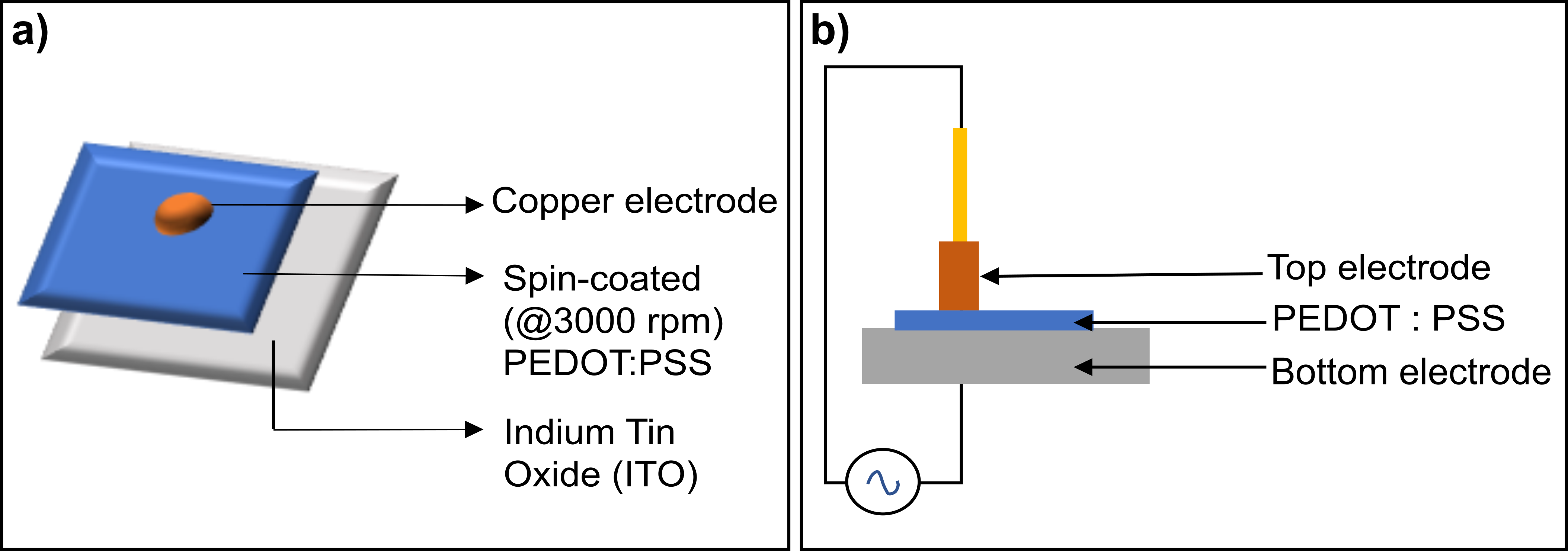}
  \caption{(a) Schematic describing polymeric thin-film device architecture. (b) and set-up for electrical characterization.}
  \label{fi:polymeric_film}
\end{figure}

Figure \ref{fi:polymeric_film}a displays the device architecture for a Cu/PEDOT:PSS/ITO device. The film is sandwiched between a copper top electrode and ITO glass bottom electrode to form a vertically stacked device. For electrical characterization of the device, voltage is directed to the top electrode via a function generator (enabling the supply of different voltage waveforms). Current is measured via the bottom electrode.

\subsection{Scheme for estimation of $V_{th}$, HRS and LRS}\label{sec:material_char}

\begin{figure}[!h]
  \centering
  \includegraphics[width=0.9\textwidth]{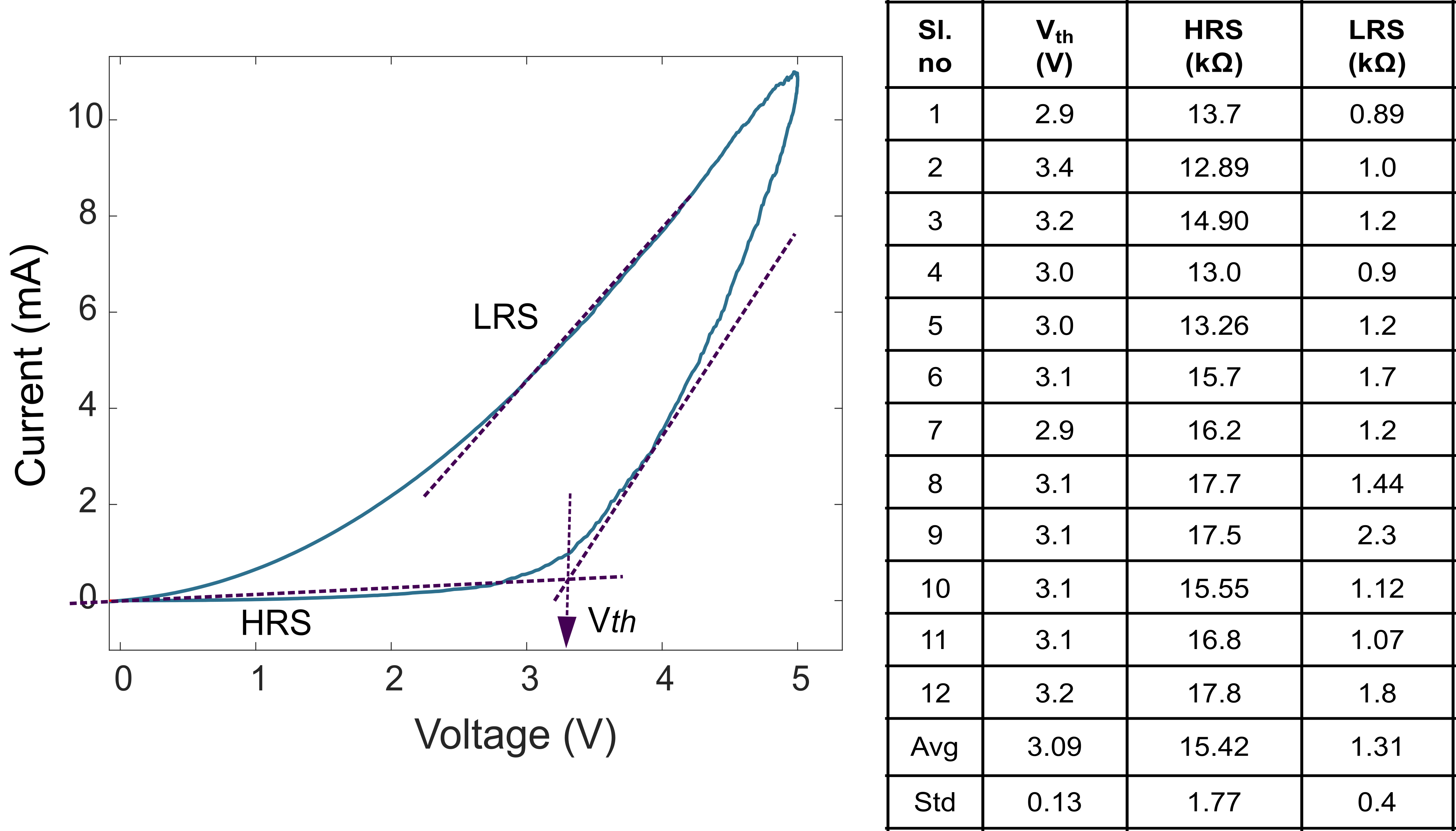}

  \caption{Technique for calculating $V_{th}$, HRS ,and LRS of PEDOT:PSS devices and the corresponding nominal values of 12 devices.}
  \label{fi:HRS_LRS}
\end{figure}

Current-voltage (i-v) characterization of the Cu/PEDOT:PSS/ITO devices reveal non-volatile hysteretic behavior with distinct switching regimes as seen in Figure \ref{fi:spatial_mem}b. Calculations were done using Matlab software to reveal the voltage threshold ($V_{th}$) of switching of resistance states of the devices as well as the estimated values of high resistance state (HRS) and low resistance state (LRS) of the devices. A conductance line of 0.033 mS was established as shown by the horizontal black dotted trace. A red dotted trace was made to fit the slope of the i-v response as the device switches to a lower resistive state. The voltage corresponding to the intersection of the red and black dotted trace highlights the voltage threshold of switching given by the blue dotted trace.
Also, to estimate the nominal values of HRS and LRS, we measured the slopes of the i-v response corresponding to the two regions as given by the brown and green trace respectively. The table displays the calculated voltage thresholds, HRS, and LRS values of 12 devices. The standard deviation reveals variability across devices within a batch which reflects device-to-device reproducibility. This highlights an important challenge in using memristive analog devices: no two devices are exactly the same.

\subsection{Repeatability of i-v behavior across cycles of periodic input}\label{sec:mem_cycles}

\begin{figure}[!h]
  \centering
  \includegraphics[width=0.65\textwidth]{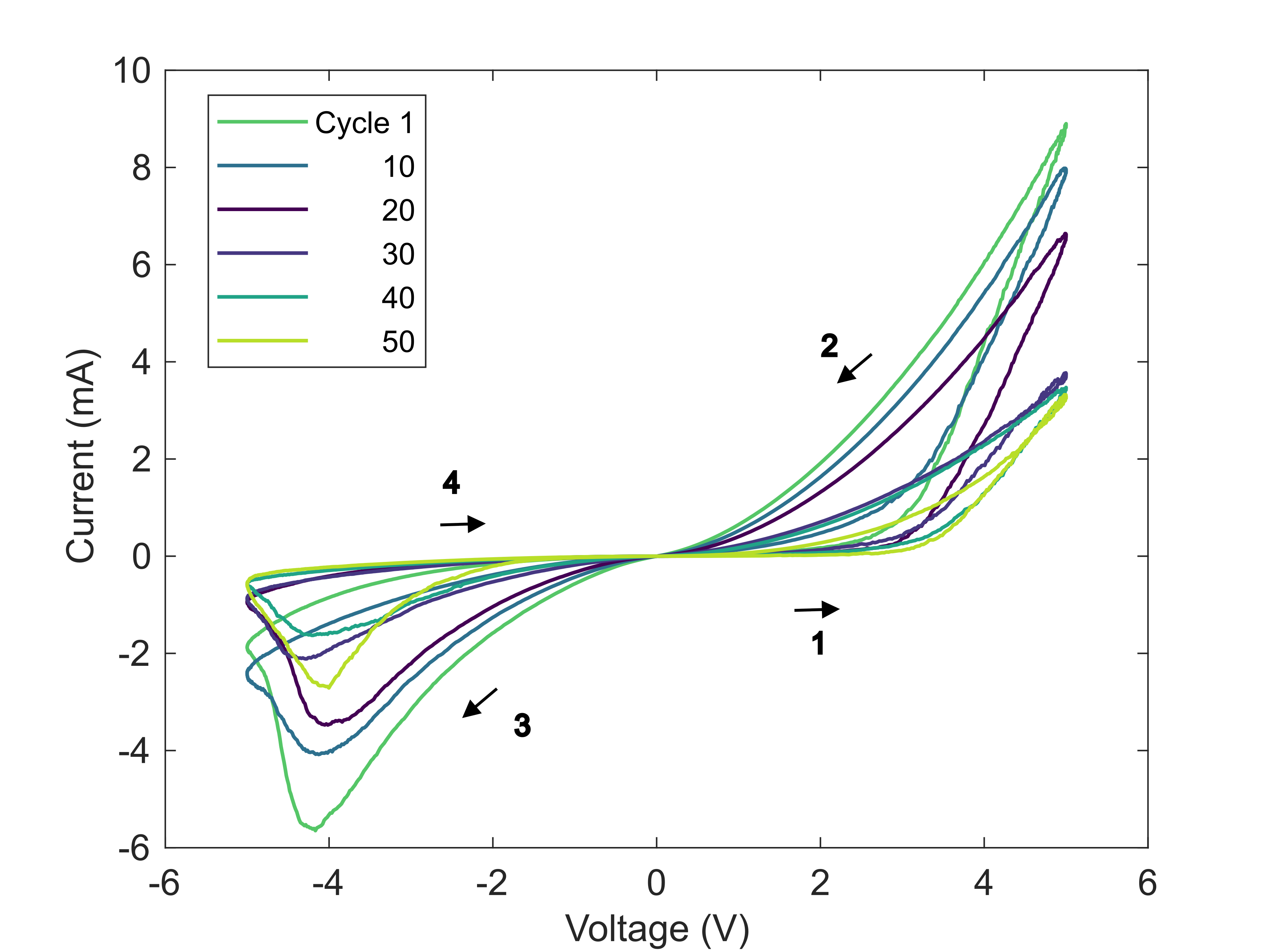}
  \caption{Cycle to cycle behavior of Cu / PEDOT:PSS / ITO device studied here.}
  \label{fi:Mem_cycles}
\end{figure}

Measurements were performed across multiple voltage sweeps to assess cycle-to-cycle variations in setting and resetting the resistive states of the devices. The former displayed more variability than the latter. Also, the slope of the i-v when the device is in the conductive domain decreases from cycle-to-cycle.

\begin{figure}[!h]
  \centering
  \includegraphics[width=0.65\textwidth]{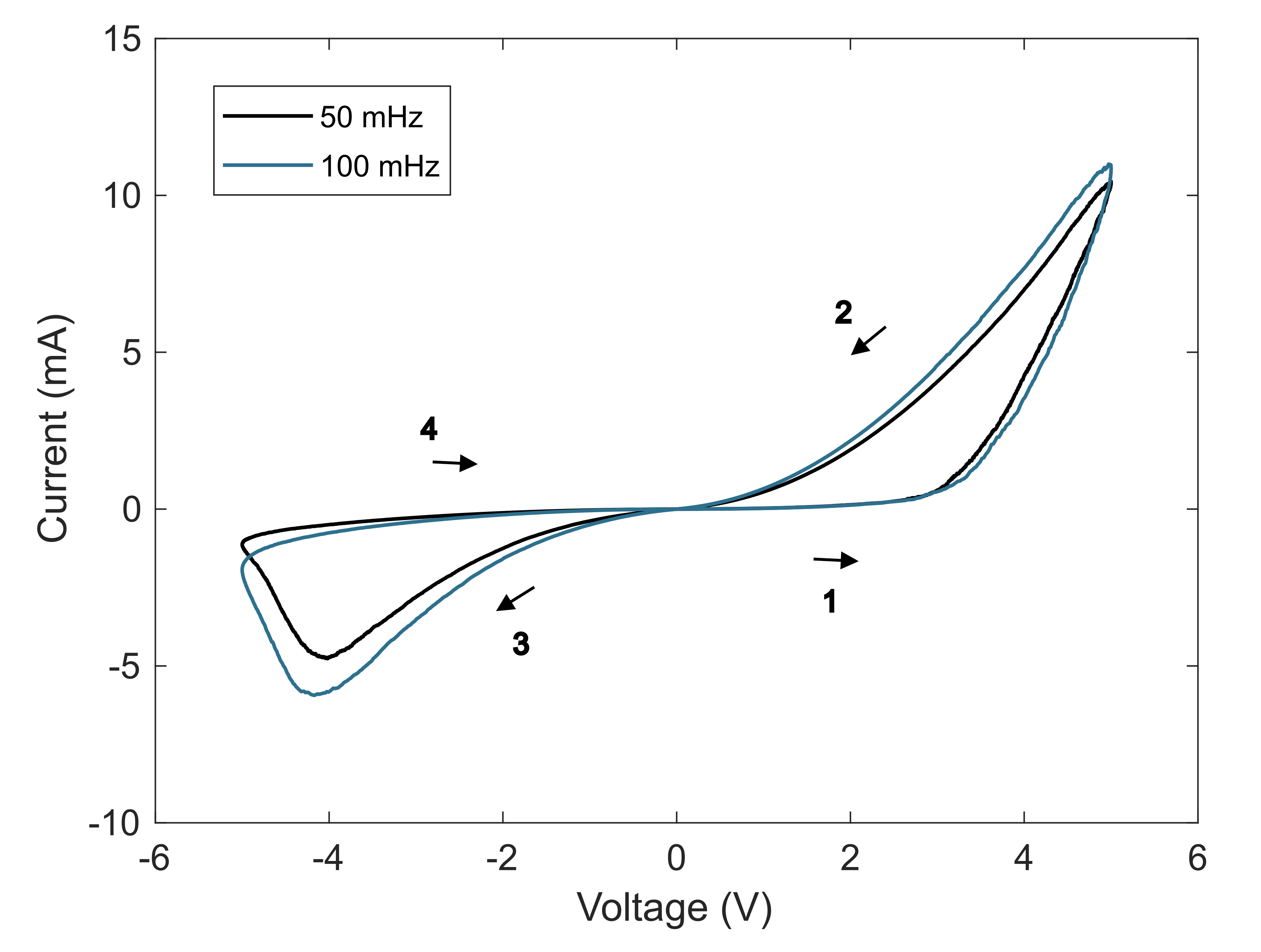}
  \caption{Response of PEDOT:PSS device to 50 mHz and 100 mHz sinusoidal voltage waveforms.}
  \label{fi:diff_frequency}
\end{figure}

When the Cu/PEDOT:PSS/ITO device is stimulated with 100 mHz and 50 mHz sinusoidal voltage stimuli, minimal change in the i-v properties of the device was noticed.

\subsection{Scheme for selection of required $V_{m}$ }\label{sec:suitable_Vm}
\begin{figure}[!h]
  \centering
  \includegraphics[width=0.65\textwidth]{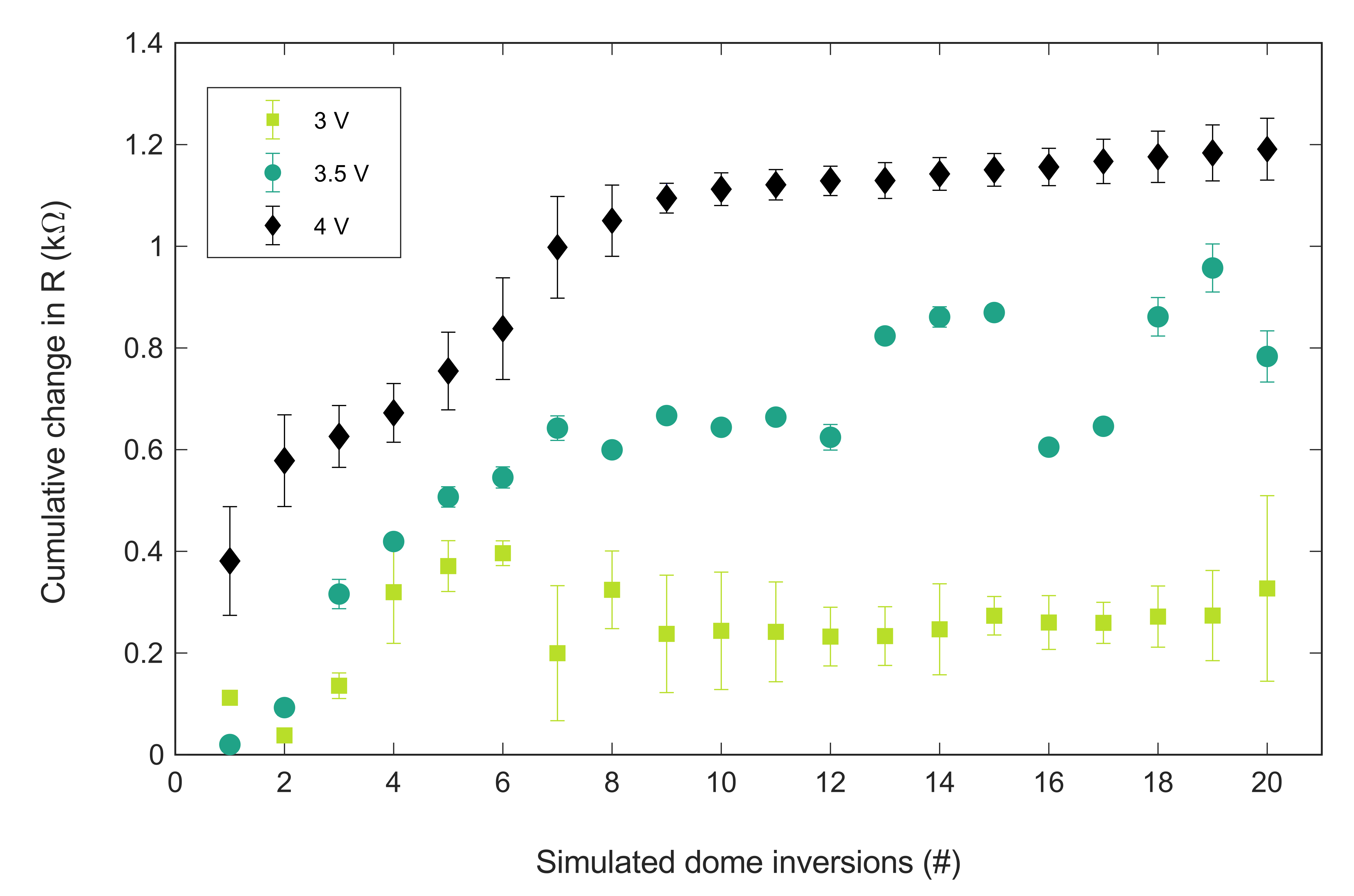}
  \caption{Cumulative change in resistance of six memristors from a higher resistive state (HRS) to a lower resistive state (LRS) across different voltages of 3,3.5 and 4V for a fixed pulse width of 5 seconds.}
  \label{fi:cum_change}
\end{figure}

To achieve a change in resistance of a memristor, the voltage across the memristor ($V_{m}$) needs to be above the $V_{th}$ of switching for the device. Figure \ref{fi:cum_change} displays the cumulative conductive response of six memristive devices to a series of pulses for various pulse amplitudes and fixed pulse width (Ton) of 5 seconds and Toff $\sim$30 seconds. The selected pulse amplitudes demonstrate the response of the memristor at $V_{th}$ (3 V), slightly beyond $V_{th}$ (3.5 V), and considerably beyond $V_{th}$ (4 V). For 3 V, we notice there is a drop in the cumulative resistance of the memristors. We also observe for pulse amplitude of 3.5 V, initially, the cumulative drop in resistance is relatively steady with respect to that of 3 V but there is significant variability or the direction of change varies significantly after several pulses. These inconsistencies motivated the usage of pulses with higher amplitude $\sim$4V, where we observe a steady cumulative decrease in resistance across 20 pulses. This comparison shows that applying a voltage well above the switching threshold improves the consistency of incremental changes in the memristive state.

\subsection{Response of device to step input and pulsatile input}\label{sec:Conductance_pulse}
\begin{figure}[!h]
  \centering
  \includegraphics[width=\textwidth]{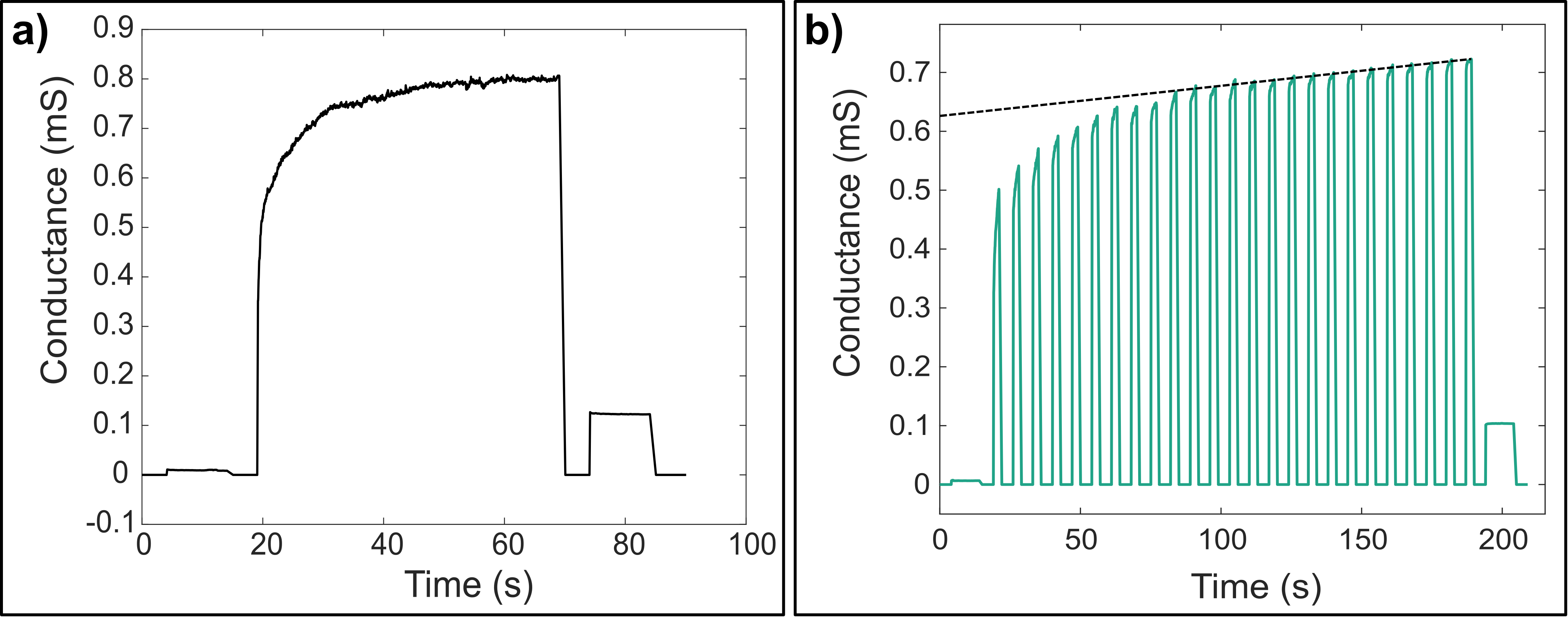}
  \caption{Response of a Cu/PEDOT:PSS/PSS device to (a) step voltage input (b) pulsatile voltage input.}
  \label{fi:Cu/PSS_Response}
\end{figure}

Response of a Cu/PEDOT:PSS/PSS device to a step voltage input of 4 V for 50 seconds revealed a significant shift in conductance from a low conductive state (HRS) to a very conductive state (LRS). The low conductive state recorded for this device was 0.8 mS which equals to 1.25 k$\Omega$, roughly around the same values for LRS recorded from i-v measurements as shown in Figure \ref{fi:Cu/PSS_Response}.  Similarly, the response of the device to consecutive pulses of amplitude 4 V and width 5 seconds revealed a cumulative increment in conductance of the device. We also observed a region of non-linear increase in conductance followed by a more linear increment in conductance indicated by the dotted green trace in Figure \ref{fi:Cu/PSS_Response}b. Therefore, a step voltage for 20 seconds can enable to shift the device conductance to a linearly incrementing domain.


\subsection{Associative Memory and Hopfield networks}\label{sec:Asso_Memory}

\begin{figure}[!h]
  \centering
  \includegraphics[width=\textwidth]{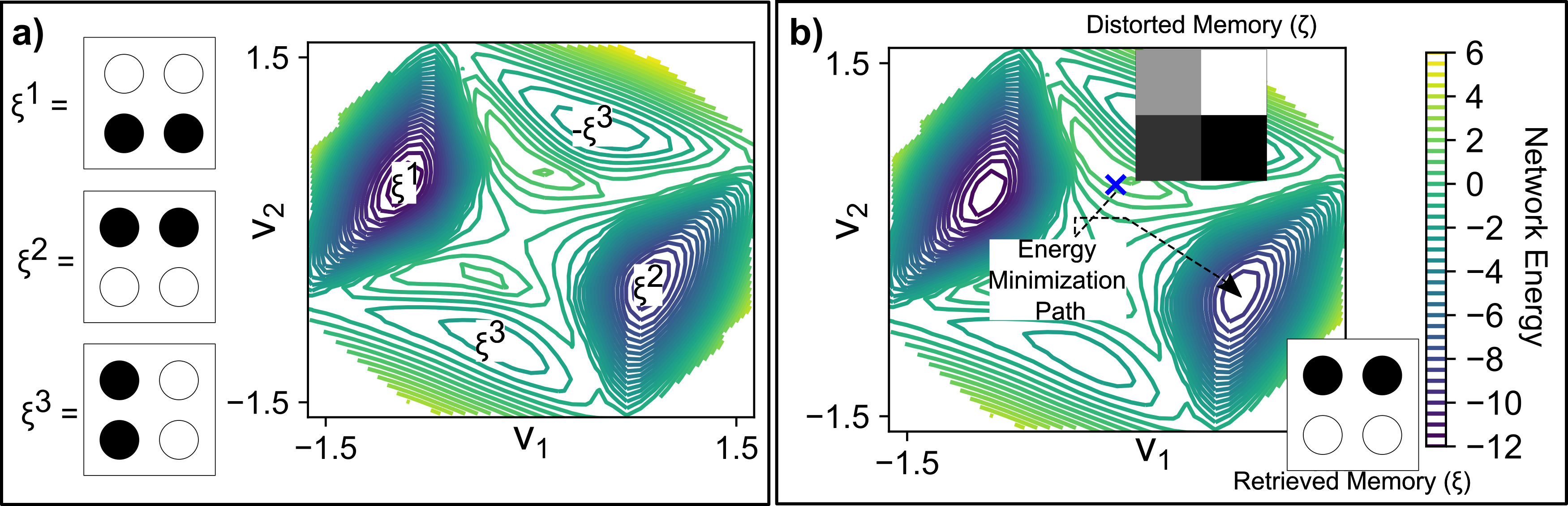}
  \caption{a) Associative memory energy landscape and stored memories b) Energy minimization path. Initial distorted pattern $\zeta$ to retrieved memory $\xi$}
  \label{fi:Asso_mem}
\end{figure}

Associative memory problems mainly focus on storing/learning a set of patterns ($\xi_i^\mu$ $\in$ $\mathbb{R}^{m\times n}$) such that when presented with new patterns ($\zeta_i$), the system responds by retrieving the closest pattern/memory \cite{Hertz2018} (Fig \ref{fi:Asso_mem}a).

Hopfield networks are a type of associative memory network which can store different memories as attractors or fixed points in their energy landscape that can then be retrieved using asynchronous neuron update and energy minimization of the network. The network is trained by computing the outer product between the patterns intended to be learned ($J_{ij}=\xi_i \xi_j$), strengthening the synapse/interaction between neurons which are both "on" (+1) or "off" (-1) and storing all the pattern information within the interaction matrix ($J_{ij}$). This same procedure can be applied to learn multiple patterns by adding the interaction matrix of each pattern as 
$J_{ij}=J_{ij}^{(1)}+J_{ij}^{(2)}+J_{ij}^{(3)}+ ..+J_{ij}^{(p)}$ \cite{Keim2019a}.
It can be observed that the training of these types of networks is just based on the strength of the interaction between neurons and a sum of the interactions matrix; hence, a physical system that can capture the additive neuron interaction over time can train the Hopfield network and store the desired patterns.
The Hopfield network mathematical model is based on the neuron state, either firing (+1) or not (-1). The state of the pattern can be calculated using the dynamics of the network \cite{Hertz2018}, as follows:
\begin{equation}\label{eq:hop_dyn}
  \xi_i^{\mu} := \text{sgn}\left(J_{ij}\xi_j^{\mu}\right)
\end{equation}
Where stored patterns are labelled by $\mu=1,2,…,p$, while the units/neurons are represented by $i=1,2,….,M$ , $\text{sgn}(x)$ is the sign function, $J_{ij}$  $\in$ $\mathbb{R}^{mn\times mn}$ is the interaction matrix between the neurons. 
Memories can be retrieved from the network using two different methods:
\begin{itemize}
    \item \textbf{Asynchronous updating:} The state of a random neuron of the input pattern $\zeta_i$ is changed (-1 to +1) independently until an energy minimum is achieved within the state $\zeta_i$. 
    
    \item \textbf{Synchronous updating:} All neurons are updated simultaneously by multiplying the interaction matrix ($J_{ij}$)  and the initial pattern ($J_{ij} \zeta_j$). This process is repeated until the input pattern is the same as the stored memory ($\xi_i=\zeta_i$). 
\end{itemize}
The energy of the network is calculated using a simplified version of the Ising model \cite{Barahona1982,Hertz2018} :

\begin{equation}\label{eq:hop_energy}
  E = -\frac{1}{2}\sum J_{ij}\zeta_i\zeta_j
\end{equation}
One can see that asynchronous updating is more suitable for any application. It does not require knowing the outcome of the network or the stored patterns, just the minimization of the network's total energy.

\subsection{Calculated interaction matrix using memristor data}\label{sec:Interaction_Matrix}

\begin{figure}[!h]
  \centering
  \includegraphics[width=0.8\textwidth]{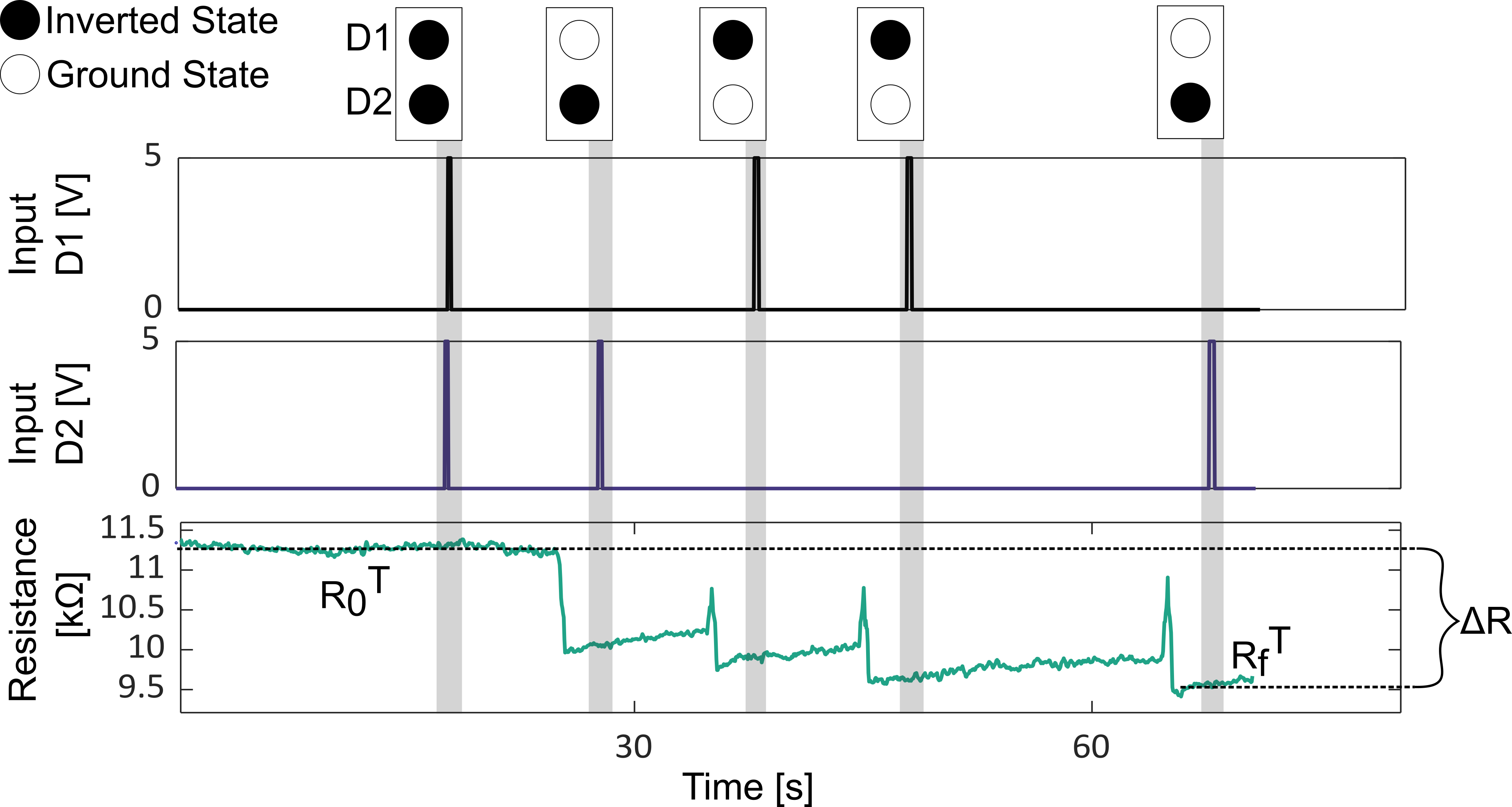}
  \caption{Interaction measurement}
  \label{fi:Mem_interac_measurment}
\end{figure}

To capture the interaction between dome units/neurons for the Hopfield network, an XOR gate connected to two different units was utilized. By doing this, the resistance of the memristor was reduced every time two units were not in the same state (see Figure \ref{fi:Mem_interac_measurment} and Movie S1). This guarantees that the values within the interaction matrix $J$ are weakened when both neurons are activated at the same time, as expected from the Hopfield network training.

The number of times two units interact with each other ($U_i$ in Figure \ref{fi:hopfield_net}) can be calculated for every memristor by taking the resistance drop. However, to reduce the noise within the memristor response and considering that $R \propto V$, this difference was calculated using voltage drop as shown in Figure \ref{fi:hopfield_net}b:

\begin{equation}\label{eq:lU_resistace}
  U_i = \text{round}\left(\frac{R_0^T-R_f^T}{\Delta R}\right)= \text{round}\left(\frac{V_0^T-V_f^T}{\Delta V}\right)
\end{equation}

Where $U_i$ is the interaction at the $i$ memristor, $R_0^T$ is the initial resistance, $R_f^T$ is the resistance value after the dome inversion events and $\Delta R$ is the average resistance change for every memristor. Experimental tests were performed by inverting the domes with the patterns shown in Figure \ref{fi:hopfield_net}c, but alternating the order within the patterns. $U_i$ results for four different experimental tests can be observed in Table \ref{tab:U_vec_comp}.

\begin{table}[!h]
\centering
\caption{$U_i$ experimental test values obtained from memristor data. Comparison with expected U vector.}
\begin{tabular}{c|cccc}\hline
Expected $U$ & Test 1 $U$ & Test 2 $U$ & Test 3 $U$ &Test 4 $U$\\\hline
1 & 1  &  1 & 1 & 1\\
2 & 2  &  4 & 2 & 2\\
3 & 3  &  3 & 3 & 3\\
3 & 3  &  3 & 3 & 3\\
2 & 0  &  1 & 1 & 1\\
1 & 1  &  0 & 1 & 0\\\hline
\end{tabular}
\label{tab:U_vec_comp}
\end{table}

$U_i$ values are arranged on the upper triangular positions of a $mn\times mn$ matrix. For a $2 \times 2$ dome array, we can write the interaction matrix as:

\begin{equation}\label{eq:interacion_matrix_mem}
J^M = 
  \begin{bmatrix}
    0 & U_1 & U_2& U_3\\
    U_1 & 0 & U_4& U_5 \\
    U_2 & U_4 & 0 & U_6 \\
    U_3 & U_5 & U_6& 0 \\
    \end{bmatrix}
\end{equation}

As memristors can only strengthen the interaction between units with opposite states, a linear transformation was stabilized to consider the weakening of units with the same state. The linear transformation for $m=n=2$ can be written as

\begin{equation}\label{eq:linear_trans_hopfield}
  \mathbf{J} = -2\mathbf{J^M} + ||\mathbf{J^M}||_{\text{max}}\mathbf{I}
\end{equation}

Where $J$ is the Hopfield network interaction matrix,  $J^M$ is the interaction matrix obtained from the memristor data, $\text{max}\left(J^M\right)$ is the maximum value within the matrix, and $I$ is a $mn\times mn$ identity matrix.

\subsection{Physical Hopfield network statistics}\label{sec:Hopfield_stat}

The physical Hopfield network and the interaction matrix performance were evaluated using the network's accuracy. This was calculated by training the physical Hopfield network with the memristors and running the energy minimization problem with 3000 corrupted patterns (see Movie S2).

\begin{figure}[!h]
  \centering
  \includegraphics[width=0.7\textwidth]{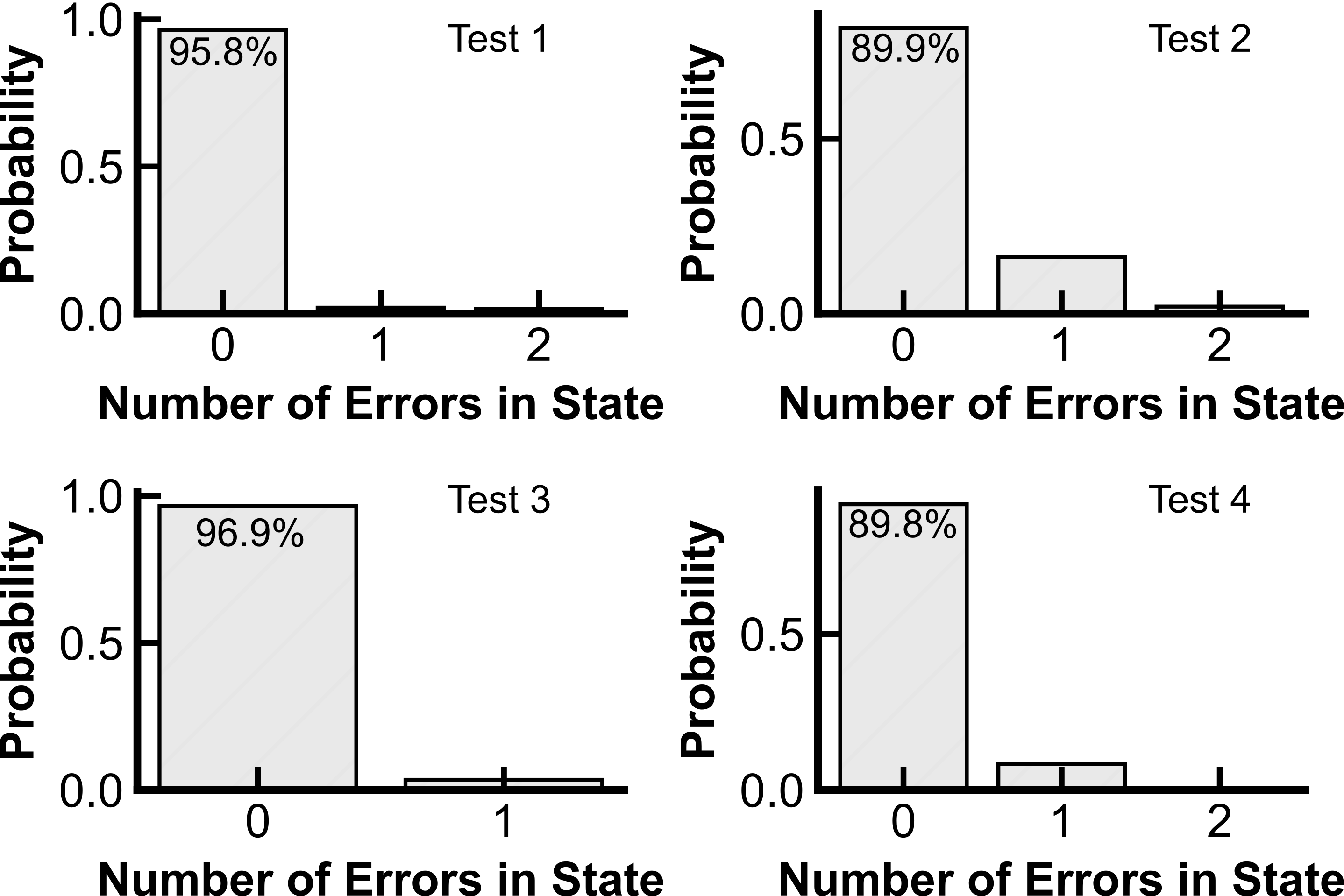}
  \caption{Probability of occurrence of errors in final retrieved memories from physical Hopfield network.}
  \label{fi:Net_Statistics}
\end{figure}

A maximum of 20 iterations and four asynchronous updates are utilized to find the stored patterns. The corrupted patterns were variations of the original ones, but with an added normally distributed noise between -1 and +1. Accuracy was calculated by counting the number of retrieved patterns that exactly match the ones on which the network was trained.  Results for the network performance for every experimental test can be observed in Table \ref{tab:net_accuracy}. 

\begin{table}[!h]
\centering
\caption{Hopfield network accuracy. Expected results with regular training and results with experimental tests.}
\begin{tabular}{c|cccc}\hline
Expected Accuracy & Test 1 & Test 2 & Test 3 &Test 4\\\hline
97.43$\%$ & 95.8$\%$  &  81.5$\%$  & 96.93$\%$  & 89.86$\%$ \\\hline
\end{tabular}
\label{tab:net_accuracy}
\end{table}

To further explore the network accuracy, we determine the probability of errors by counting the number of state errors in the retrieve patterns and the number of times the error was present. Results from all experimental tests can be observed in Figure \ref{fi:Net_Statistics}.

\end{document}